\documentclass[twocolumn]{aastex63}
\usepackage[version=4]{mhchem}
\usepackage{graphicx}
\usepackage{multirow}
\newcommand\sigmadet{\sigma_{\mathrm{det}}}

\shortauthors{Currie et al.}

\begin{document}

\title{There’s more to life than \ce{O2}: Simulating the detectability of a range of molecules for ground-based high-resolution spectroscopy of transiting terrestrial exoplanets}

\correspondingauthor{Miles Currie}
\email{mcurr@uw.edu}

\author[0000-0003-3429-4142]{Miles H. Currie}
\affiliation{Department of Astronomy and Astrobiology Program, University of Washington, Box 351580, Seattle, Washington 98195, USA}
\affiliation{NASA Nexus for Exoplanet System Science, Virtual Planetary Laboratory Team, Box 351580, University of Washington, Seattle, Washington 98195, USA}

\author[0000-0002-1386-1710]{Victoria S. Meadows}
\affiliation{Department of Astronomy and Astrobiology Program, University of Washington, Box 351580, Seattle, Washington 98195, USA}
\affiliation{NASA Nexus for Exoplanet System Science, Virtual Planetary Laboratory Team, Box 351580, University of Washington, Seattle, Washington 98195, USA}
\affiliation{Astrobiology Center, 2-21-1 Osawa, Mitaka, Tokyo 181-8588, Japan}

\author[0000-0002-0470-0800]{Kaitlin C. Rasmussen}
\affiliation{Department of Astronomy and Astrobiology Program, University of Washington, Box 351580, Seattle, Washington 98195, USA}
\affiliation{NASA Nexus for Exoplanet System Science, Virtual Planetary Laboratory Team, Box 351580, University of Washington, Seattle, Washington 98195, USA}

\begin{abstract}

Within the next decade, atmospheric \ce{O2} on Earth-like M dwarf planets may be accessible with visible--near-infrared, high spectral resolution extremely large ground-based telescope (ELT) instruments. However, the prospects for using ELTs to detect environmental properties that provide context for \ce{O2} have not been thoroughly explored.  Additional molecules may help indicate planetary habitability, rule out abiotically generated \ce{O2}, or reveal alternative biosignatures.  To understand the accessibility of environmental context using ELT spectra, we simulate high-resolution transit transmission spectra of previously-generated evolved terrestrial atmospheres. We consider inhabited pre-industrial and Archean Earth-like atmospheres, and lifeless worlds with abiotic \ce{O2} buildup from \ce{CO2} and \ce{H2O} photolysis. All atmospheres are self-consistent with M2V--M8V dwarf host stars. Our simulations include explicit treatment of systematic and telluric effects to model high-resolution spectra for GMT, TMT, and E-ELT configurations for systems 5 and 12 pc from Earth. Using the cross-correlation technique, we determine the detectability of major species in these atmospheres: \ce{O2}, \ce{O3}, \ce{CH4}, \ce{CO2}, \ce{CO}, \ce{H2O}, and \ce{C2H6}. Our results suggest that \ce{CH4} and \ce{CO2} are the most accessible molecules for terrestrial planets transiting a range of M dwarf hosts using an E-ELT, TMT, or GMT sized telescope, and that the \ce{O2} NIR and \ce{H2O} 0.9 $\mu$m bands may also be accessible with more observation time. Although this technique still faces considerable challenges, the ELTs will provide access to the atmospheres of terrestrial planets transiting earlier-type M-dwarf hosts that may not be possible using JWST.

\end{abstract}


\section{Introduction}\label{sec:intro}

With next-generation space- and ground-based observatories capable of characterizing the atmospheres of terrestrial exoplanets coming online, we enter a new era of searching for signs of habitability and life. For the best targets, including the TRAPPIST-1 transiting planet system, spectral characterization with the James Webb Space Telescope (JWST) may reveal novel environments that contain signs of planetary evolution influenced by abiotic or possibly biological processes \citep[e.g.][]{Lincowski2018-zc, Lustig-Yaeger2019-bk, Wunderlich2019, Pidhorodetska2020-gv, Gialluca2021-cq, Meadows2023}. The upcoming extremely-large ground-based telescopes (ELTs) will provide an additional near-term opportunity to understand how Earth-like planets evolve under a variety of conditions, and search for signs of life. In particular, the ELTs have the potential to complement the capabilities of space-based telescopes by accessing \ce{O2} and other molecules that can give context to \ce{O2}. While \ce{O2}, the major byproduct of oxygenic photosynthesis, and \ce{O3}, the photochemical byproduct of \ce{O2}, are unlikely to be detectable with JWST \citep{Lustig-Yaeger2019-bk, Wunderlich2019, Pidhorodetska2020-gv, Meadows2023}, a large body of previous work suggests that \ce{O2} may be detectable from the ground \citep{Rodler2014, Kawahara2012-cw, Serindag2019, Lopez-Morales2019, Snellen2013}. The next generation of ground-based telescopes and their instruments have several advantages for \ce{O2} detection, including the ability to spectrally resolve narrow features at shorter wavelengths, where \ce{O2} more strongly absorbs, and a broader range of M dwarf targets \citep{Snellen2013, Rodler2014, Lopez-Morales2019}. Consequently, ground-based observations will likely be highly complementary to the near- and mid-infrared studies of terrestrial exoplanet atmospheres possible with  JWST.

The feasibility of detecting \ce{O2} in ground-based observations of terrestrial exoplanets is relatively well studied. The main challenge of this approach is separating exoplanet \ce{O2} absorption from Earth's \ce{O2} absorption; however, it has been argued that with well-constrained radial velocity measurements from high-resolution (R $\sim$ 100,000) ELT spectrometers \citep{Szentgyorgyi2014-xj, Marconi2022-sr, Mawet2019}, this challenge can be overcome. Using this method, simulations suggest that strong atmospheric \ce{O2} features may be detected through the Earth's atmosphere by observing less than 40 transits for nearby targets \citep{Snellen2013, Rodler2014, Serindag2019, Lopez-Morales2019}, making \ce{O2} a prime biosignature gas to be sought by upcoming ground-based telescopes. \citet{Snellen2013} pioneered the use of cross-correlation analysis to simulate how to untangle the telluric oxygen lines from exoplanet oxygen, and calculated that \ce{O2} is detectable at 3.8$\sigma$ with the European--ELT (E-ELT) in 30 transits for a hypothetical Earth twin in the GJ 1214 (M5V) system $\sim14.6$ pc away. \citet{Rodler2014} later refined this study by adding red noise and atmospheric refraction effects to the simulated observations, and predicted about twice as many transits would be needed to detect \ce{O2} at the same significance. Incorporating the effects of more realistic noise, \citet{Serindag2019} injected an \ce{O2} signal into observations of Proxima Centauri using the Ultraviolet and Visual Echelle Spectrograph (UVES) instrument on the Very Large Telescope (VLT), and found that \ce{O2} would be detectable in 20--50 transits for an Earth twin orbiting an M5 star 7 pc away, which is consistent with both  \citet{Snellen2013} and \citet{Rodler2014}. \citet{Lopez-Morales2019} sought to optimize instrumentation and observing strategies for detecting \ce{O2}, and found that increasing the spectral resolution to $R=300,000$ more than doubles the \ce{O2} line depths compared to $R=100,000$ spectra, and the higher spectral resolution lowers the number of transits to detect \ce{O2} at 3$\sigma$ by 34\%.

While the ground-based high-resolution technique is well explored for \ce{O2} detection, similar investigations for the detectability of molecules that can help interpret \ce{O2} in the context of its planetary environment are relatively sparse. \citet{Lovis2017} explored \ce{O2}, \ce{CH4}, and \ce{H2O} detectability for planetary reflected light observations using the SPHERE+ESPRESSO instrumentation on the VLT and found that \ce{O2} and \ce{H2O} may be accessible on an Earth-like Proxima Centauri b, but the unavailability of high-accuracy line list data in their wavelength region limited their ability estimate \ce{CH4} detectability. More recently, \citet{Lin2020} suggested that in addition to \ce{O2}, ELTs may be used to search for potential biosignature or climate-indicator gases like \ce{CH4}, \ce{CO2}, and \ce{H2O} in Earth-like atmospheres around late-type M dwarfs, but did not simulate observations to quantify molecular detectability for high-resolution spectra. \citet{Wunderlich2020-fv} adopt a simple approach to estimate the detectability of a suite of molecules including \ce{CO2}, \ce{CH4}, \ce{O2}, \ce{H2O}, and \ce{CO} using an ELT, and finds that \ce{CO2} and \ce{CH4} may be readily accessible for TRAPPIST-1 e, but a rigorous cross-correlation analysis of simulated observations was outside of the scope of their paper.

While \ce{O2} is undoubtedly an important biosignature for life on Earth, a detection of even abundant \ce{O2} alone in an exoplanet atmosphere is not necessarily a sign of life.  Confidence that abundant \ce{O2} is in fact due to life can be increased if abiotic mechanisms for its generation (so-called ``false positives'') can be ruled out \citep{Meadows2017-wx}. Post ocean-loss worlds, for example, can generate extremely abundant (up to thousands of bars), potentially detectable levels of \ce{O2} on planets orbiting M dwarfs, and may be a common outcome of atmospheric evolution \citep{Luger2015-ys, Tian2015-ak, Schwieterman2016-mg, Lustig-Yaeger2019-bk}, but could potentially be discriminated in high-resolution spectroscopic observations via detection of shorter-wave \ce{O2} bands in addition to a non-detection of the 1.27$\mu$m infrared band \citep{Leung2020}.  \ce{CO2} photolysis could also be a source of abiotic \ce{O2} in circumstances that inhibit \ce{CO} and \ce{O} recombination \citep{Domagal-Goldman2014-oh, Tian2014-fs, Harman2015-oc, Gao2015-ck, Hu2012-jr}, including inhibiting NO$_\mathrm{x}$ production \citep{Harman2018-gq}, in which case the concurrent presence of \ce{CH4}, which is often low in cases of abiotic \ce{O2} production \citep{Domagal-Goldman2014-oh}, or \ce{CO}, the other byproduct of \ce{CO2} photolysis, may be false positive discriminants. The interpretation of an \ce{O2} detection, therefore, is highly dependent on other characteristics of the planetary environment, and will require the detection of other gases or bands to help provide this context.

When modeling M dwarf planets, which are better suited to study in transmission, it is also important to account for star-planet interactions that can alter the atmosphere of habitable zone planets away from an Earth-twin scenario. On M dwarf planets, even with the same assumed surface fluxes of gases, the spectral energy distribution (SED) of the host star can alter molecular destruction and production rates.  These in turn alter photochemical lifetimes of gas species, potentially enhancing the abundance of key gases in the atmosphere such as \ce{CH4} \citep{Segura2005-mg}.  Planets in the habitable zone may also receive different irradiances than Earth depending on their orbital geometries \citep{Kopparapu2013-wm} which may act to cool or warm the planet relative to Earth.  However, the carbonate--silicate cycle, which is presumed to be active on a planet with liquid water and plate tectonics, may stabilize the surface temperature by buffering the \ce{CO2} abundance in the planetary atmosphere \citep{Walker1981-wj}.  Thus planets receiving less irradiance from their host star may have a buildup of \ce{CO2} in their atmospheres to keep the surface temperate. These different abundances and distributions of atmospheric species due to the action of photochemistry and climate buffers can significantly alter their detectability, and thus these effects must be accounted for when interpreting observations of terrestrial exoplanet environments. 

Expanding detectability predictions to include molecules other than \ce{O2} has distinct advantages in helping us search for additional biosignatures and for more comprehensively assessing any biosignatures that we may find, including \ce{O2}.  
The likelihood that the presence of \ce{O2} is a biosignature needs to be assessed by searching for molecules that help us characterize the underlying planetary environment and its history, rule out false positives for any biosignatures detected, and potentially strengthen the biosignature interpretation in other ways by identifying disequilibria that are more likely to be driven by life \citep{Meadows2018-yx,Krissansen-Totton2016-bq}.  In particular, the detection of a strong \ce{CO2} feature immediately increases the chance that a given planet is terrestrial, but may also suggest a possible source for abiotic \ce{O2}.  Determining whether observed \ce{O2} is more or less likely to be due to an \ce{O2}-dominated atmosphere generated by \ce{H2O} photolysis from past ocean loss can potentially be achieved by searching at multiple wavelengths for the presence of strong collisionally-induced absorption from \ce{O2} molecules in a dense atmosphere   \citep{Lincowski2018-zc,Lustig-Yaeger2019-bk,Leung2020}.  However, there are several other possible false positive discriminant molecules that can be used to indicate than an abiotic source may be more likely, such as the simultaneous presence of \ce{CO2} and \ce{CO}, its photolytic byproduct  \citep{Schwieterman2016-mg}, and whether or not water vapor is present  \citep{Wordsworth&Pierrehumbert2014, Gao2015-ck}. As well as ruling out abiotic sources, additional molecules, especially in disequilibrium pairs, can make a stronger case for a biological source.  Examples include the canonical \ce{O2}/\ce{CH4} pair \citep{Hitchcock1967-ky} that would reveal the gigayear-long impact of oxygenic photosynthesis on our planet \citep{Lyons2014-yw}, coupled with high fluxes of biogenic \ce{CH4} generated by methanogens, that greatly exceed anticipated geological production rates \citep[][]{Etiope2013-fg}. Even in more exotic planetary environments, it is challenging to produce high abiotic \ce{CH4} fluxes without also revealing the abiotic mechanism through other observable features \citep{Thompson2022-nd}.  Finally, by expanding the types of molecules sought, we can also look for alternative biosignatures that are potentially more detectable, such as the \ce{CH4}/\ce{CO2} disequilibrium pair \citep{Krissansen-Totton2016-bq} that may have been present and potentially detectable for the majority of the Earth's history \citep{Meadows2023}. This biosignature could also be made more robust by additionally searching for and ruling out the presence of CO, which, if present, may indicate enhanced outgassing of abiotic \ce{CH4} from a more reduced interior \cite{Krissansen-Totton2016-bq,Krissansen-Totton2018-tj}. 

In this work, we explore the potential for using the ELTs to detect a number of molecules expected in transiting terrestrial planetary atmospheres, and their application for biosignature interpretation.  We use previously-generated simulations of self-consistent inhabited environments for Earth throughout its history transiting five M dwarf spectral types from M2V-M8V.  We apply cross-correlation techniques to simulated spectra of these environments to determine the number of transits needed to detect the molecules. Additionally, we explore two lifeless, but habitable, atmosphere types that may also exhibit abiotically-generated \ce{O2} from processes such as significant \ce{CO2} photolysis, complete ocean loss, and ongoing ocean outgassing, to understand how we can potentially discriminate these atmospheres from those influenced by life.  In Section~\ref{sec:methods}, we outline how we constructed synthetic observed spectra and cross-correlated them with a template spectrum to estimate the detectability of molecular bands in the visible and near-infrared. In Section~\ref{sec:results}, we present the results of our detectability calculations, which inform future observations of transiting terrestrial exoplanets. In Section~\ref{sec:discussion}, we discuss the significance of detections and non-detections of molecular bands in the context of habitability and life, and propose an observational strategy for characterizing terrestrial atmospheres in transit. We summarize our findings in Section~\ref{sec:conclusion}.

\section{Methods}\label{sec:methods}

\begin{figure*}
    \centering
    \includegraphics[width=0.75\textwidth]{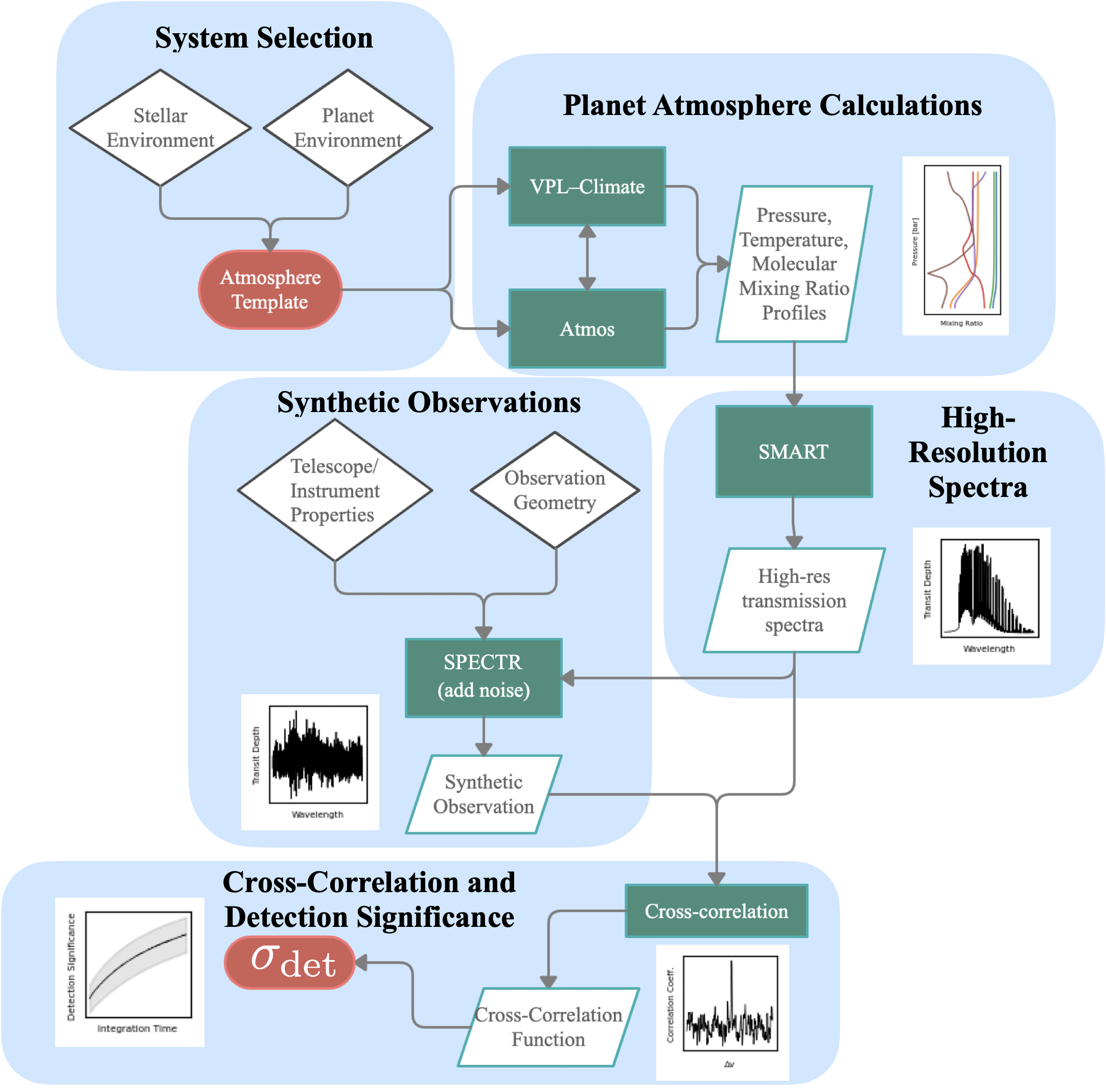}
    \caption{Overview of methods pipeline. The red round boxes represent the start and end points of the pipeline. The solid green boxes represent models or software pipelines, the white parallelograms represent input or output data products, and the rhombuses represent user-specified metadata or parameters. Cartoon examples of the data products are shown next to each data product node.}
    \label{fig:pipeline}
\end{figure*}

In this work, we develop a novel pipeline for estimating the detectability of molecular bands in transiting terrestrial exoplanet atmospheres using cross-correlation analysis (Figure~\ref{fig:pipeline}). The model takes simulated high-resolution (R=100,000) transit transmission spectra of a variety of terrestrial environments as its input, adds appropriate noise for ground-based high-resolution spectroscopy, and performs a cross-correlation analysis on the synthetic observation and a model spectrum. We use the resulting cross-correlation function to determine the detectability of a molecular band as a function of the number of transits observed.

\subsection{Model inputs}
Our novel detectability pipeline (Figure~\ref{fig:pipeline}) requires high-resolution transit transmission spectra as its input. We generate high-resolution spectra using a radiative transfer model applied to atmospheric profiles that describe the pressure, temperature and mixing ratio of key gases as a function of altitude (pressure). To generate atmospheres that are self-consistent with the host star SEDs, we use results from previous coupled climate--photochemical models. 

\subsubsection{System selection}
We test the detectability of multiple absorption bands for seven molecules (Table~\ref{tab:bands}) in four atmosphere classes derived from the climate/photochemistry models of other studies  (Table~\ref{tab:experiments}) orbiting five different M dwarf host stars, at 5 and 12 pc away from Earth. To inform the development of future instrumentation, the detectability is calculated for individual absorption bands of each molecule, as \citet{Lopez-Morales2019} found no advantage to combining molecular bands in observations when accounting for red noise.  The four atmosphere classes include pre-industrial Earth (PIE) and Archean Earth (ARE) \citep{Meadows2018-yx, Davis2022inprep}, a scenario with abiotic \ce{O2} buildup from \ce{CO2} photolysis (referred to as \ce{CO2} photolysis) for both lightning on and lightning off cases  \citep{Harman2018-gq}, and scenarios for 10 bar \ce{O2} atmospheres that result from both complete ocean loss (referred to as ocean-loss) with internal desiccation, and ongoing ocean loss with subsequent interior outgassing (referred to as ocean outgassing) \citep{Meadows2018-yx,Leung2020}. All planets are one Earth radius in size. We place the  PIE and ARE planets in orbit around M2V, M3V, M4V, M6V, and M8V dwarf host stars, and use outputs from previous studies that produced self-consistent photochemical and climatic simulations for calculating our spectra. The \ce{CO2} photolysis and 10 bar \ce{O2} ocean loss/outgassing atmospheres are self-consistent with only M4V and M6V dwarf hosts, respectively, and thus detectability for these planets orbiting other M dwarf hosts is not considered. For our pre-industrial Earth-like atmospheres, we include both clear sky and cloudy scenarios.  In total, there are $14$ unique planet/star systems in this study, outlined in Table~\ref{tab:experiments}. As an example of a potential real ELT target, we calculate the detectability of species in our Earth-sized PIE and ARE atmospheres with TRAPPIST-1 e's orbit (M8V host) at its canonical distance of 12 pc \citep{Gillon2016-et} using the atmosphere models of Davis et al. (in prep).  

\begin{deluxetable}{cc}\label{tab:bands}
\tablecaption{Molecular bands explored in this study}
\tablewidth{0pt}
\tablehead{
\colhead{Molecule} &  \colhead{Bands [$\mu m$]}
}
\startdata
O$_2$ & 0.69, 0.76, 1.27 \\
CH$_4$ & 0.89, 1.1, 1.3, 1.6  \\
CO$_2$ & 1.59, 2.0 \\
H$_2$O & 0.94, 1.1, 1.3 \\
O$_3$ & 0.63, 0.65, 3.2 \\
CO & 1.55, 2.3 \\
C$_2$H$_6$ & 3.33 \\
\enddata
\end{deluxetable}

\begin{deluxetable*}{ccc}\label{tab:experiments}
\tablecaption{Atmosphere classes and host stars}
\tablewidth{0pt}
\tablehead{
\colhead{Atmosphere} &  \colhead{Host Star(s)} & \colhead{Atm. Ref.}
}
\startdata
Pre-industrial Earth (PIE) & M2V, M3V, M4V, M6V, M8V & Davis et al. (in prep) \\
Archean Earth (ARE) & M2V, M3V, M4V, M6V, M8V & Davis et al. (in prep) \\
\ce{CO2} photolysis, lightning ON & M4V & \citet{Harman2018-gq} \\
\ce{CO2} photolysis, lightning OFF & M4V & \citet{Harman2018-gq} \\
10 bar \ce{O2} complete ocean loss & M6V & \citet{Meadows2018-yx} \\
10 bar \ce{O2} ongoing ocean outgassing & M6V & \citet{Meadows2018-yx} \\
\enddata
\end{deluxetable*}

\begin{deluxetable*}{cccccc}
\tablecaption{Planet orbital properties\label{tab:planets}}
\tablewidth{0pt}
\tablehead{
\colhead{Host Spectral Type} & \colhead{Transit duration [hr]} & \colhead{Time between transits [hr]} & \colhead{Orbital Period [day]} & \colhead{Semi-major axis [AU]}
}

\startdata
M2 & 4.7 & 1153 & 48 & 0.24 \\
M3 & 3.6 & 793 & 33 & 0.19 \\
M4 & 3.4 & 692 & 29 & 0.16 \\
M6 & 1.2  & 152 & 6.4 & 0.041 \\
M8 & 0.99 & 96 & 4.0 & 0.027 \\
\enddata

\end{deluxetable*}

\begin{deluxetable*}{ccccccccc}
\tabletypesize{\footnotesize}
\tablewidth{0.99\textwidth}
\tablecaption{Host star properties\label{tab:stars}}
\tablehead{
\colhead{Spectral Type} & \colhead{Example star} & \colhead{Radius } & Synthetic Spectrum Ref. & m$_{I}$ (5 pc) & m$_{J}$ (5 pc) & m$_{I}$ (12 pc) & m$_{J}$ (12 pc) 
\\
 & & [R$_\odot$] & & [mag] & [mag] & [mag] & [mag}
\startdata
M2 & GJ 832  & 0.499$^{a}$ & \citet{Peacock2019-ol} & 6.2 & 5.0 & 8.1 & 6.9 \\ % \pm 0.017$$^{2}$ \\
M3 & GJ 436  & 0.464$^{b}$ &  \citet{Peacock2019-ol} & 6.8 & 5.4 & 8.7 & 7.3 \\
M4 & GJ 876 &  0.3761$^{c}$ & \citet{Peacock2020-cc} & 7.1 & 5.8 & 9.0 & 7.7 \\
M6 & Proxima Centauri & 0.141$^{d}$ & Davis et al. (in prep) & 10.7 & 8.6 & 12.6 & 10.5  \\ %, 0.1410 +/-  0.0070$^{7}$  \\
M8 & TRAPPIST-1 & 0.114$^{e}$ &  \citet{Peacock2019-bz} & 12.2 & 9.3 & 14.1 & 11.2\\% +/- 0.006$^{10}$   \\
\enddata
\tablecomments{Stellar radii references:
$^a$\citet{Houdebine2010-cu},
$^b$\citet{Torres2007-cz},
$^c$\citet{Von_Braun2014-xs},
$^d$\citet{Bonfils2005-pb},
$^e$\citet{Filippazzo2015-wv}. The magnitudes are calculated for each example star placed at 5 and 12 pc away from Earth. }

\end{deluxetable*}

\subsubsection{Planet atmosphere calculations}
For the self-consistent planetary atmospheres used as input to the simulator, we use previously generated results from a 1-D coupled climate--photochemistry model---which includes the effects of the host star SED on the photochemistry and climate of each planet atmosphere. The model couples the atmospheric chemistry component of \textit{Atmos}, a publicly available model based on \cite{Kasting1979-tx} and \citet{Zahnle2006-pk}, with VPL-Climate, a general-purpose, 1D radiative--convective equilibrium, terrestrial planet climate model
\citep{Meadows2018-yx, Robinson2018-ja}. The coupled model is described in detail in \citet{Lincowski2018-zc}, including validations for both Earth and Venus. 

\begin{figure*}
\centering
    \includegraphics{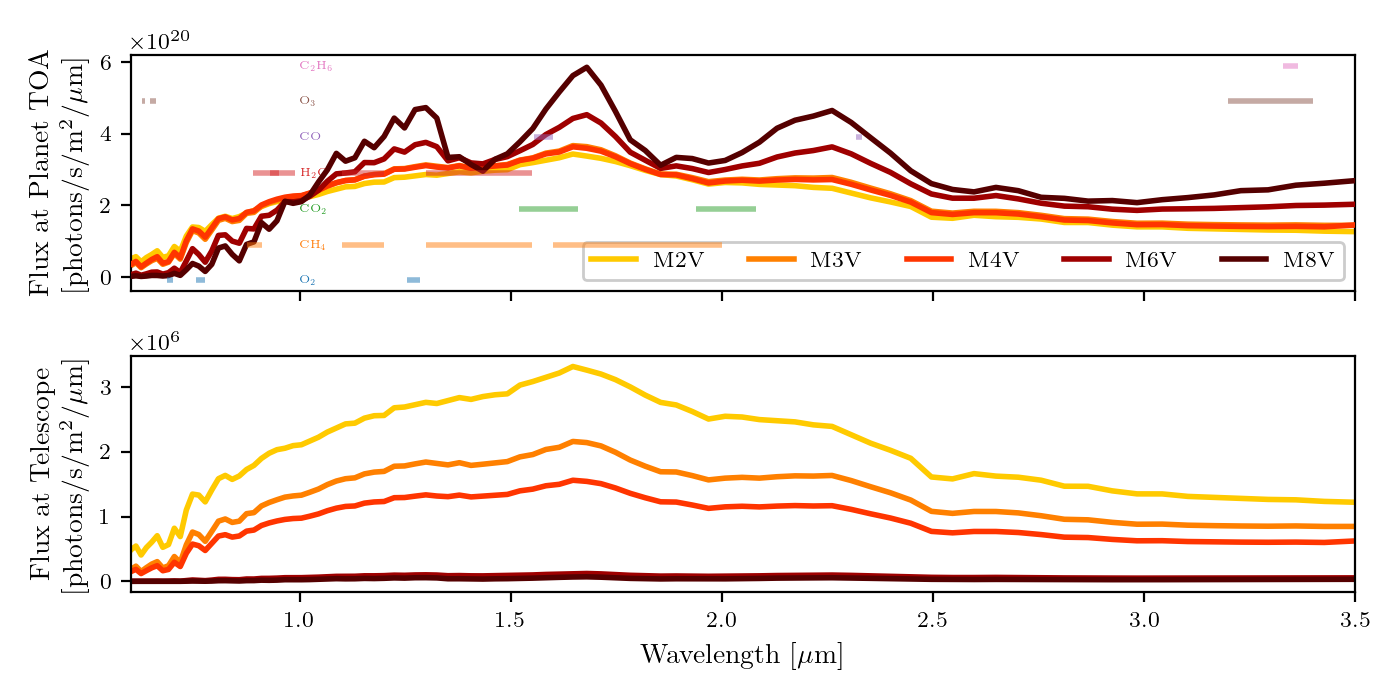}
    \caption{\textbf{Upper panel:} Spectral energy distributions of the M dwarf host stars in this study incident at the top of the planets' atmospheres. \textbf{Lower panel:} The stellar fluxes as seen by an observer 12 pc away. The spectral resolution is binned down to $R=50$ for clarity. The colored horizontal lines are selected molecular absorption bands we investigate in our planetary atmospheres.}
    \label{fig:stellar}
\end{figure*}

\begin{figure*}
\centering
    \includegraphics[width=\textwidth]{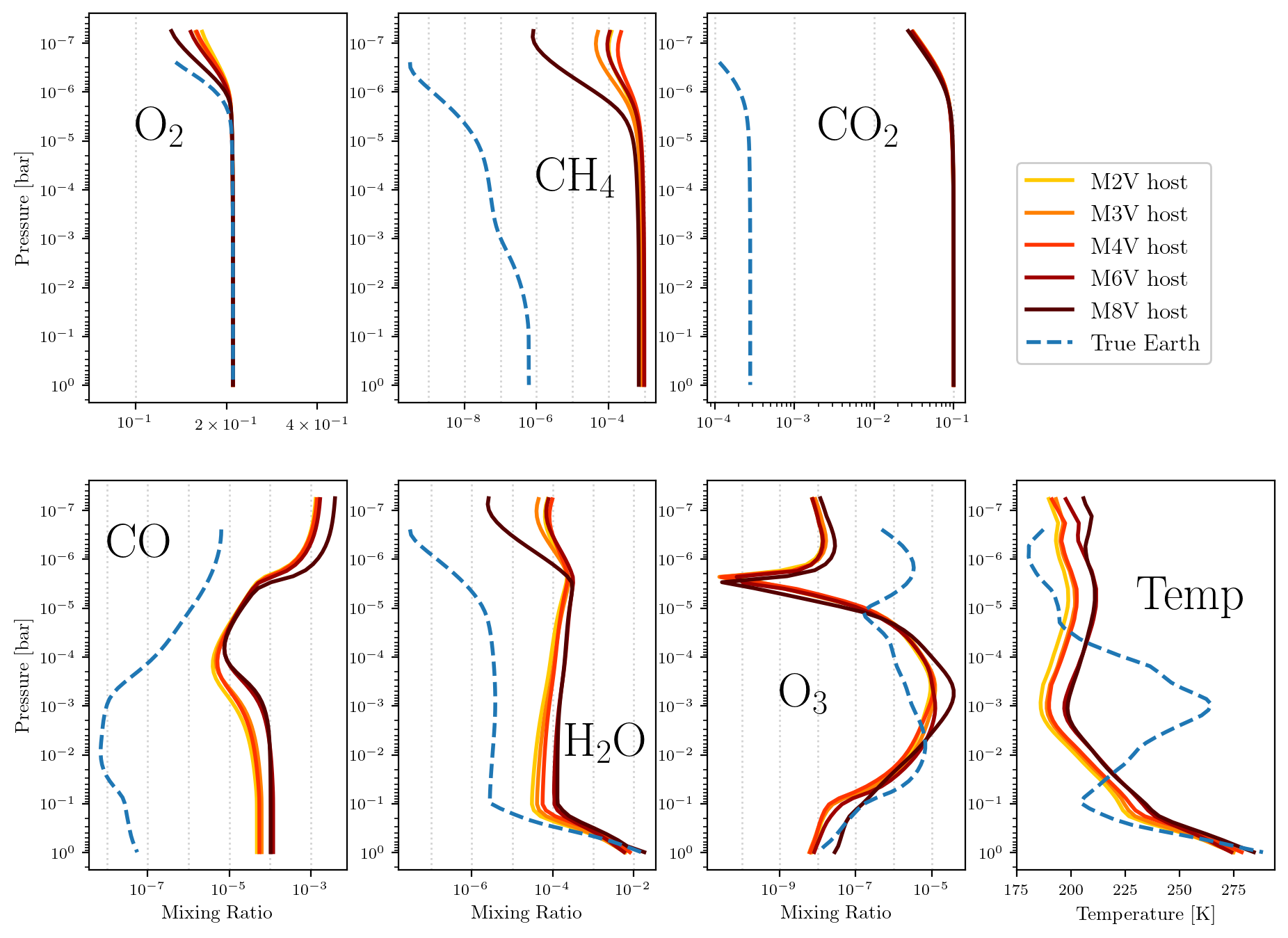}
    \caption{Mixing ratios for the major species in the pre-industrial Earth-like (PIE) atmospheres and temperature profiles. Each line represents a self-consistent Earth-like atmosphere orbiting a stellar host. Earlier-type M dwarf hosts are lighter and later-type M dwarf hosts are darker. The dashed blue line represents profiles for the Earth orbiting the Sun for comparison.  }
    \label{fig:spaghettipie}
\end{figure*}

\begin{figure*}
\centering
    \includegraphics[width=\textwidth]{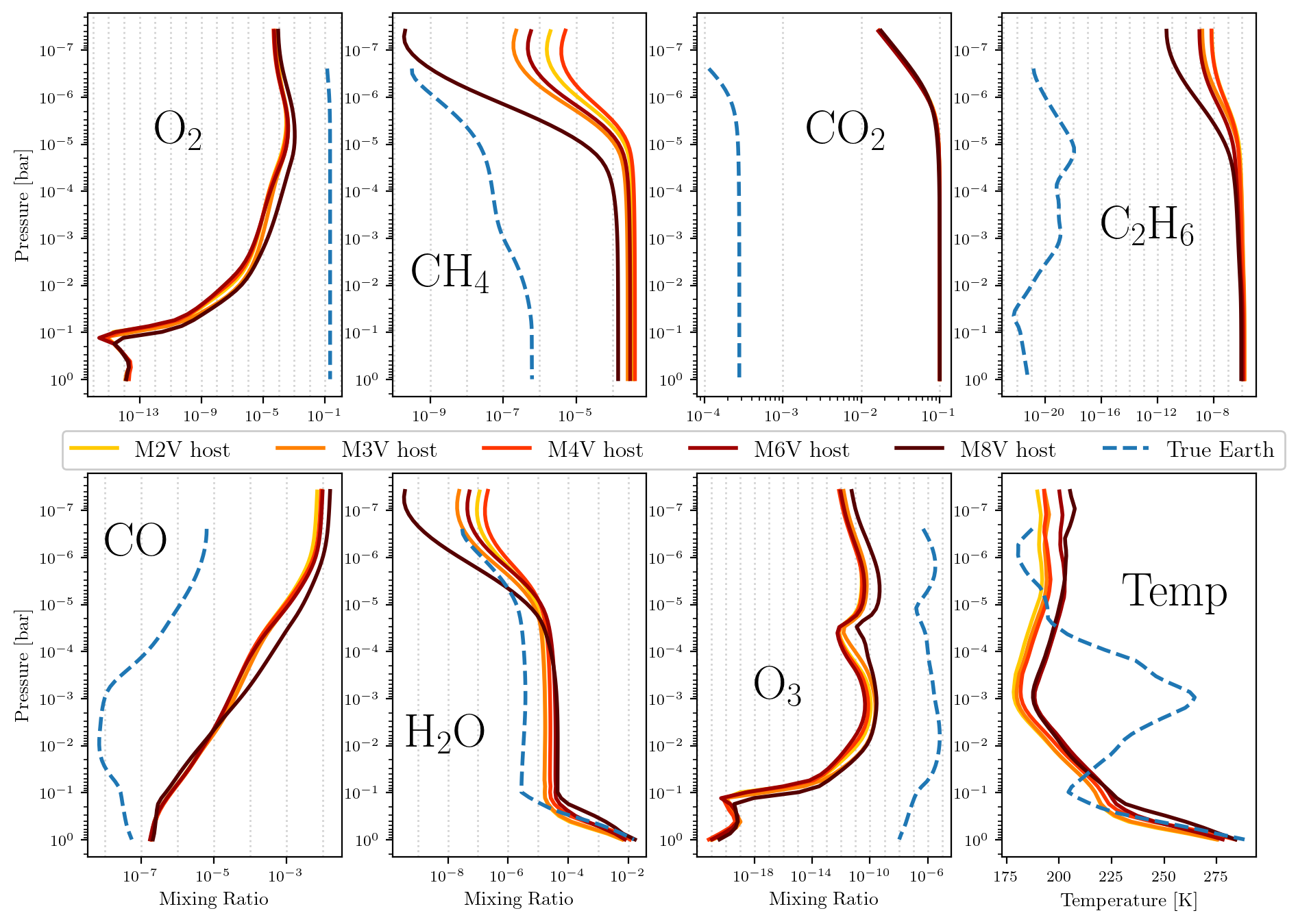}
    \caption{Mixing ratios for the major species in the Archean Earth-like (ARE) atmospheres and temperature profiles. Each line represents a self-consistent Earth-like atmosphere orbiting a stellar host. Earlier-type M dwarf hosts are lighter and later-type M dwarf hosts are darker. The dashed blue line represents profiles for the Earth orbiting the Sun for comparison.  }
    \label{fig:spaghettiarch}
\end{figure*}

Davis et al. (in prep) updated the coupled climate--photochemistry model to include new \ce{H2O} cross-sections, and ran globally-averaged PIE and ARE atmospheres orbiting M2V, M3V, M4V, M6V, and M8V dwarf stars to convergence, which we use for our PIE and ARE atmosphere classes in this study. Each planet is placed in orbit around its host star such that it receives 0.66 times the irradiance that Earth receives, and the accompanying orbital properties are given in Table~\ref{tab:planets}.  This irradiance is approximately the same as that received by the M dwarf habitable zone planets TRAPPIST-1e \citep{Gillon2017-jf} and Proxima Centauri b \citep{Anglada-Escude2016-ai}.  The host star properties and spectra are given in Table~\ref{tab:stars}, and their synthetic spectra are plotted in Figure~\ref{fig:stellar} (degraded to R = 50 for clarity). For each spectral type, we use publicly available synthetic high-resolution stellar spectra for the stars GJ832, GJ436, GJ876, Proxima Centauri, and TRAPPIST-1 (\citet{Peacock2019-bz, Peacock2019-ol, Peacock2020-cc}, Davis et al. in prep.).  These spectra serve as analog examples for M2V, M3V, M4V, M6V, and M8V dwarf stars, respectively.
The vertical gas abundance and temperature profiles for the self-consistent PIE and ARE atmosphere calculations are shown as solid lines in Figures~\ref{fig:spaghettipie} and~\ref{fig:spaghettiarch}, respectively; we include true Earth profiles in dashed lines as a comparison. The habitable PIE and ARE atmospheres considered here contain 10\% \ce{CO2}, which is required to raise the global mean surface temperatures above the freezing point of water \citet{Meadows2018-yx}.  This abundance is not unreasonable if we assume that the surface temperature of a planet within the habitable zone is buffered by a carbonate-silicate feedback, as proposed by \citep{Walker1981-wj}.  With the carbonate-silicate feedback working, a planet may have an atmospheric \ce{CO2} abundance of up to several bars of \ce{CO2} near the HZ outer edge \citep{Kopparapu2013-wm}.

\begin{figure*}
\centering
    \includegraphics[width=\textwidth]{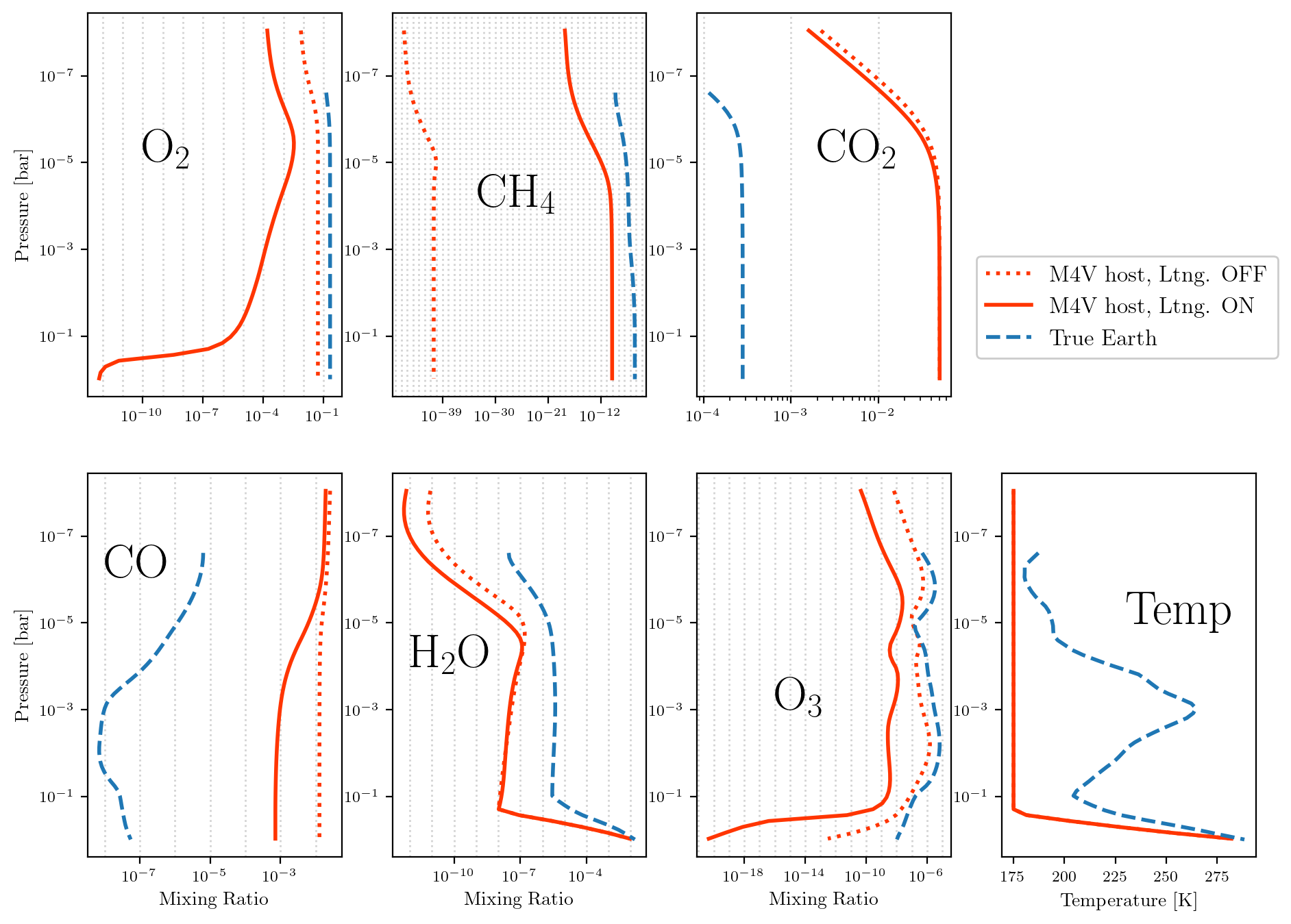}
    \caption{Mixing ratios for the major species in the abiotic \ce{O2} buildup from \ce{CO2} photolysis atmospheres and temperature profiles. The solid and dotted red lines represent profiles for a \ce{CO2 photolysis} atmosphere with lightning turned on and off, respectively, orbiting an M4V-type stellar host. The dashed blue line represents profiles for the Earth orbiting the Sun for comparison. }
    \label{fig:spaghettisonny}
\end{figure*}

The \citet{Harman2018-gq} \ce{CO2} photolysis atmospheres, which, with lightning turned off, can generate up to 6\% \ce{O2} from \ce{CO2} photolysis are self-consistent with our M4V stellar host, and the corresponding molecular and temperature profiles are presented in Figure~\ref{fig:spaghettisonny}. The two red lines show the effects of turning lightning off (dotted) and on (solid). \ce{NO} produced by lightning catalyzes the recombination of \ce{CO} and \ce{O}, decreasing oxidizing species and increasing reducing species in the atmospheres. Again, true Earth is shown for comparison in the blue dashed lines.

\begin{figure*}
\centering
    \includegraphics[width=0.7\textwidth]{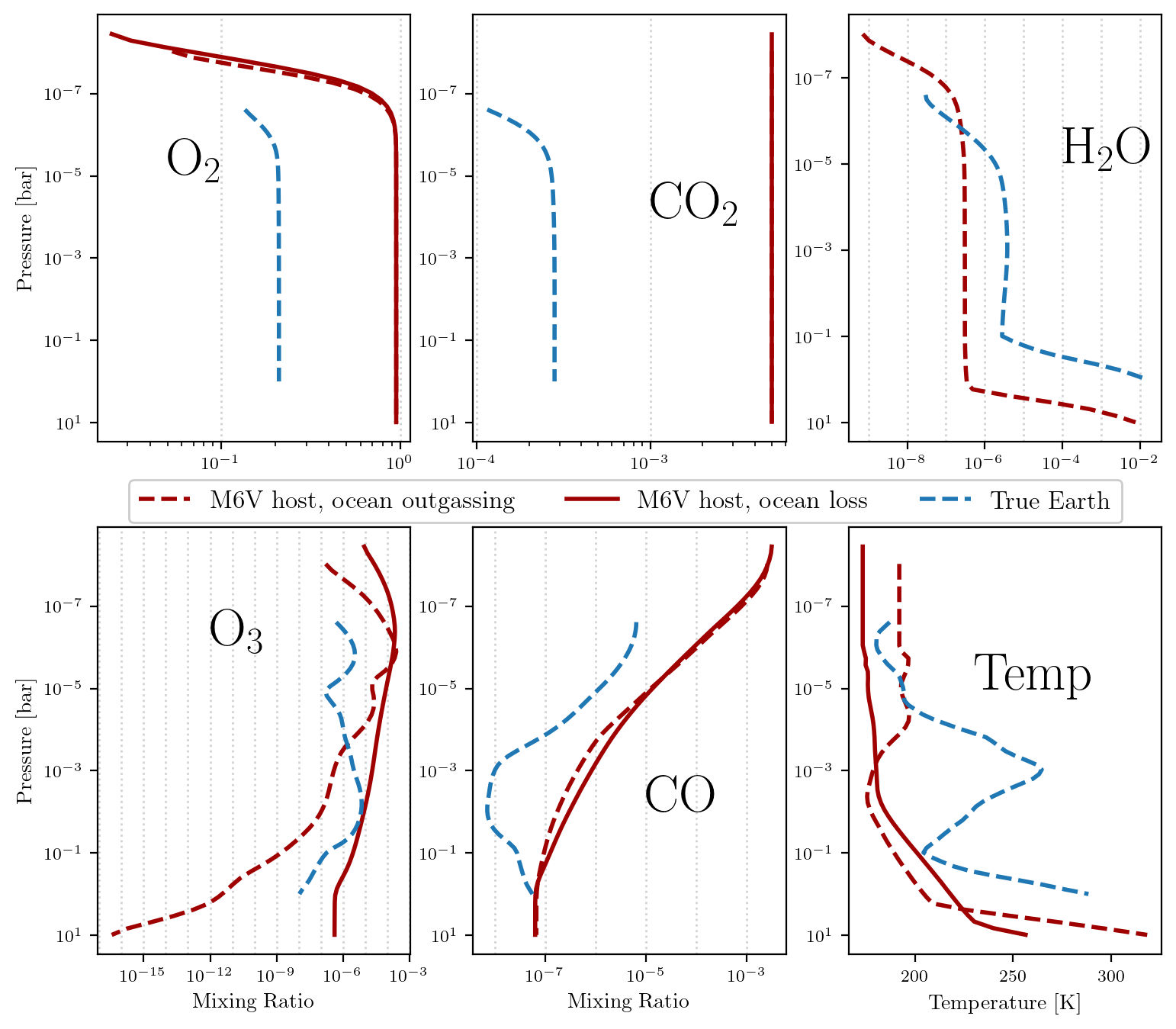}
    \caption{Mixing ratios for the major species in the 10-bar \ce{O2} ocean outgassing (dashed) and ocean-loss (solid) atmospheres and temperature profiles. The dashed blue line represents profiles for the Earth orbiting the Sun for comparison. There is no \ce{H2O} in the ocean-loss atmosphere. }
    \label{fig:spaghettimichaela}
\end{figure*}

The 10-bar \ce{O2} atmospheres from \citet{Leung2020} are self-consistent with our M6V stellar host, and the corresponding molecular and temperature profiles are presented in Figure~\ref{fig:spaghettimichaela}, with ocean-loss profiles as the solid lines and ocean-outgassing profiles as the dashed lines. True Earth profiles are shown for comparison as the blue dashed lines.

\begin{figure*}
    \centering   
    \includegraphics[width=\textwidth]{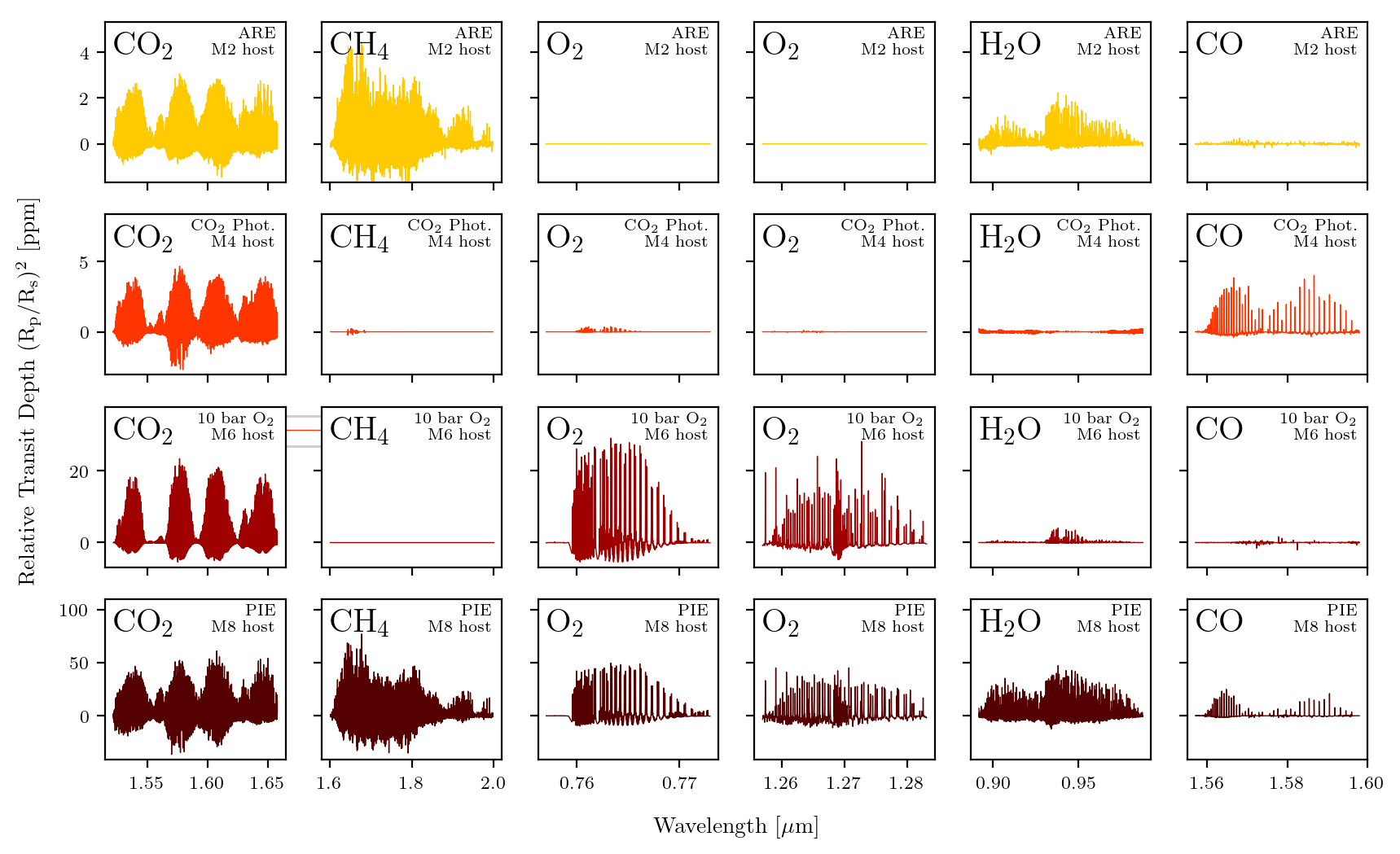}
    \caption{Selected molecular bands in this work. Each molecular band is plotted at $R=100,000$ and is continuum subtracted using a high-pass filter (see Section~\ref{sec:methods}). Note the scale in each row: planets transiting later type M dwarf hosts typically have larger relative transit depths than planets transiting early type hosts.}
    \label{fig:bands}
\end{figure*}

\begin{figure*}
\centering
    \includegraphics[width=\textwidth]{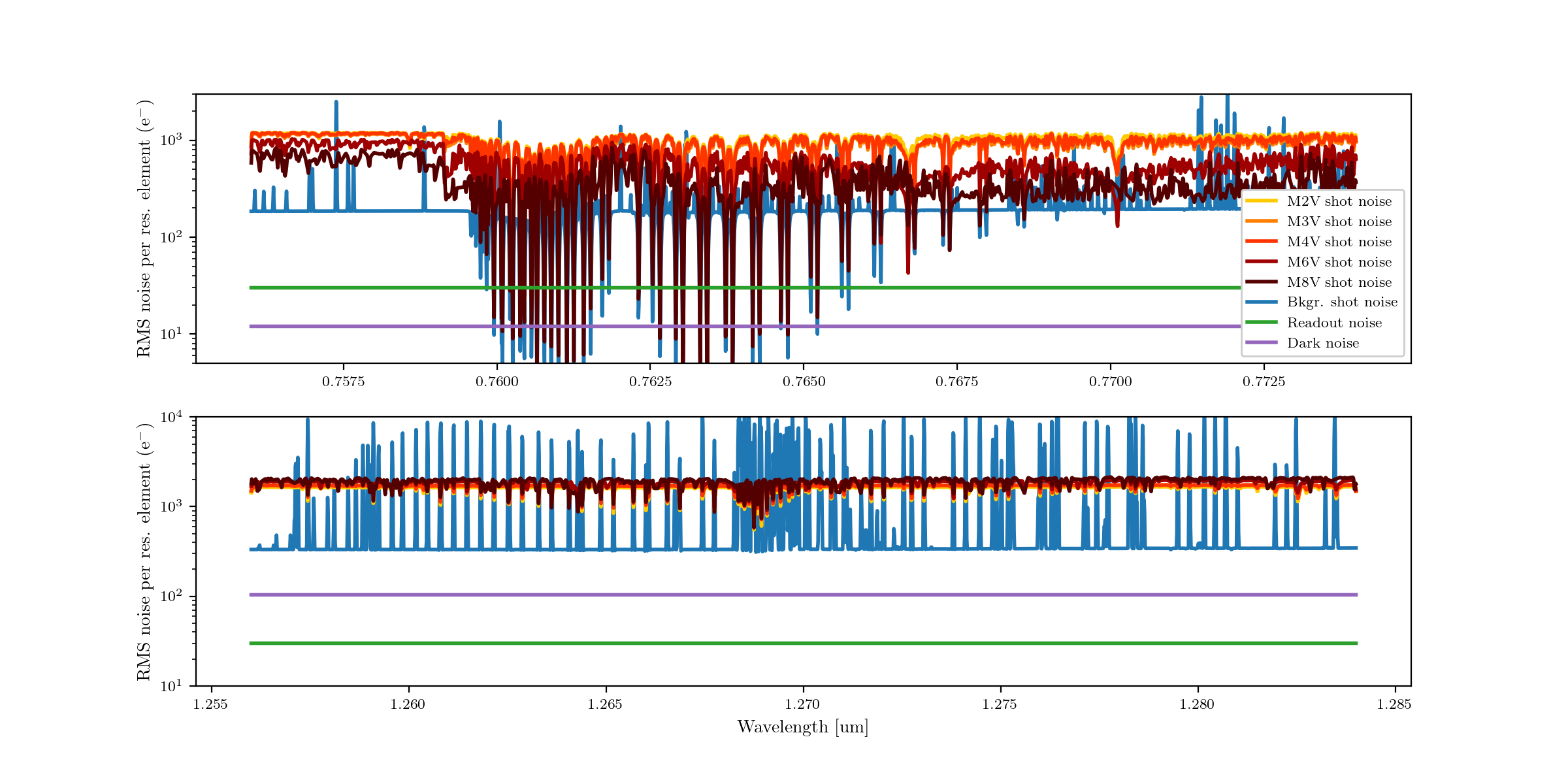}
    \caption{Noise sources per resolution element considered in this work for a band in the visible (top panel) and a band in the NIR (bottom panel) for a 2 hr exposure at R = 100,000. The background shot noise is comprised of zodiacal light, telluric emission, airglow, and telescope/instrument thermal emission, and the airglow continuum dominates the background continuum. The shot noise is plotted for all stars at 12 pc, and includes attenuation from the Earth's telluric transmission. Photon (shot) noise from the M2V and M8V stars exceeds the combined readout and dark noise by a factor of 25 and 8, respectively, for the visible, and 12 and 15, respectively, for the NIR.}
    \label{fig:noise}
\end{figure*}

\subsubsection{High-resolution spectra}
We calculated high-resolution spectra of our transiting planet atmospheres using SMART, a 1D, line-by-line radiative transfer model \citep{Stamnes1988,Meadows1996-fp, Crisp1997-fn}. SMART calculates the spectra using stellar sources, and solves the radiative transfer equation for each atmospheric constituent in discrete layers in the atmosphere. In each layer, SMART calculates extinction due to vibrational and rotational transitions, and collisionally-induced absorption for each absorbing gas. It also calculates the effects of aerosols, Rayleigh scattering, and wavelength-dependent surface albedo. SMART outputs top-of-atmosphere planetary radiances and transmission spectra, and has been validated for Earth in reflected light at low-resolution in \citet{Robinson2011-os}, and for Earth in transmission at high-resolution in \citet{Lustig-Yaeger2022}.  We calculated each spectrum at a resolution of $R=1,000,000$ for both clear and cloudy sky scenarios, which we then convolve with a Gaussian profile to achieve a resolution of $R=100,000$ for this study. As in \citet{Lopez-Morales2019}, we find that increasing the resolution of our spectra to $R=300,000$ roughly doubles (100\% increase) the average line depths for the \ce{O2} A-band and NIR band. However, this effect is dependent on the molecular band, and we see different increases in line depth ranging from  50\% (O3, 0.65 um) to nearly 200\% (CO, 2.3 um).  We simulate cloudy-sky scenarios as 50\% clear sky, 25\% Earth-like cirrus clouds, and 25\% and Earth-like stratocumulus clouds for the pre-industrial Earth-like atmospheres. The clouds are assumed to be uniformly-spaced around the planet, and the cirrus and stratocumulus clouds are placed at 0.331 bar and 0.847 bar, respectively. Figure~\ref{fig:bands} show clear-sky high-resolution spectra for selected molecular bands in this study.

\subsection{Simulated observations}
We upgraded an existing, sophisticated noise model, \texttt{coronagraph}---which is designed for simulating observations of space-based telescopes \citep{Robinson2016-lw, Lustig2019coronagraph}---and modified it  to produce synthetic observations of ground-based ELTs. Because we are simulating transmission spectra for the ELTs, we turn off the coronagraph mode in our ground-based upgrade of \texttt{coronagraph}. To avoid confusion, we hereafter refer to our coronagraph-free version of the \texttt{coronagraph} code as the Spectral Planetary ELT Calculator for Terrestrial Retrieval (SPECTR) pipeline\footnote{https://github.com/curriem/spectr}.  To estimate the wavelength-dependent signal-to-noise ratio for a ground-based  observation, we use SPECTR to calculate the incoming photon count from the exoplanet/star system, and the background photon count from  zodiacal, exo-zodiacal, telescope, instrument, detector, and atmospheric (telluric) sources. 

\subsubsection{Modeling the Sky}
To simulate ground-based observations, SPECTR has a built-in interface to the Cerro Paranal Advanced Sky Model \citep[SykCalc][]{Noll2012-br, Jones2013-ex}, a highly customizable telluric atmosphere model built by the European Southern Observatory (ESO) for observation planning. Within SPECTR, the user can choose parameters that describe the observatory site, season, time, target coordinates, and moon phase/location, which are passed to the SkyCalc command line interface. SkyCalc returns the wavelength-dependent background and telluric lines at the native resolution of the instrument. The background component is comprised of moonlight, starlight, zodiacal light, telluric emission, airglow, and telescope/instrument thermal emission. We find that the dominant component of the background is the airglow continuum, which is comprised of radiation from chemiluminsecent reactions between atmospheric constituents \citep[e.g.][]{Khomich2008, Kenner1984} as well as pseudo-continua from many closely spaced molecular lines \citep[e.g.][]{Saran2011}.  We use Paranal as our observatory site, with precipitable water vapor of 3.5 mm, and do not include scattered moonlight in these simulations.

\subsubsection{Telescope and instrument properties}\label{sec:telinstprop}

We simulate $R = 100,000$ observations for European ELT (E-ELT), Thirty Meter Telescope (TMT), and Giant Magellan Telescope (GMT) configurations, with collecting areas of and 978 m$^2$, 707 m$^2$, and 368 m$^2$, respectively. These collecting areas correspond to a 39 m E-ELT with a central hole to accommodate its secondary mirror\footnote{https://elt.eso.org/about/facts/}, a 30 m diameter TMT, and the total expected collecting area of all mirror segments of the GMT\footnote{https://giantmagellan.org/explore-the-design/}. We expect our E-ELT aperture to yield a 1.39x higher flux than our TMT aperture. The required equivalent observation time with our E-ELT configuration is $\sim30$\% less than our TMT configuration to yield the same flux. The E-ELT aperture yields a 2.66x higher flux than our GMT aperture, and the required equivalent E-ELT observation time is $\sim60$\% less than for the GMT. The ELTs will be equipped with high-resolution spectrometers capable of R $=$ 100,000 in visible and/or near-infrared wavelengths with estimated throughputs of 10\%, typical dark current values of 0.0002 \ce{e-}/pix/s and 0.015 \ce{e-}/pix/s for the visible and NIR, respectively, and read noise of 3 \ce{e-}/pix \citep[GMT]{Szentgyorgyi2014-xj} \citep[E-ELT]{Marconi2022-sr} \citep[TMT]{Mawet2019}. We use these values for our E-ELT, TMT, and GMT configurations for this study, and assume 100 detector pixels per resolution element. In particular, we note that the term $T$ in Equation~\ref{eq:cr} is the 10\% throughput multiplied by the telluric absorption. One can approximate the effect of varying the total efficiency by multiplying our detection significance calculation by the square root of the ratio of a new efficiency and the 10\% efficiency we use in our calculations. 

\begin{figure}
    \centering
    \includegraphics[width=0.45\textwidth]{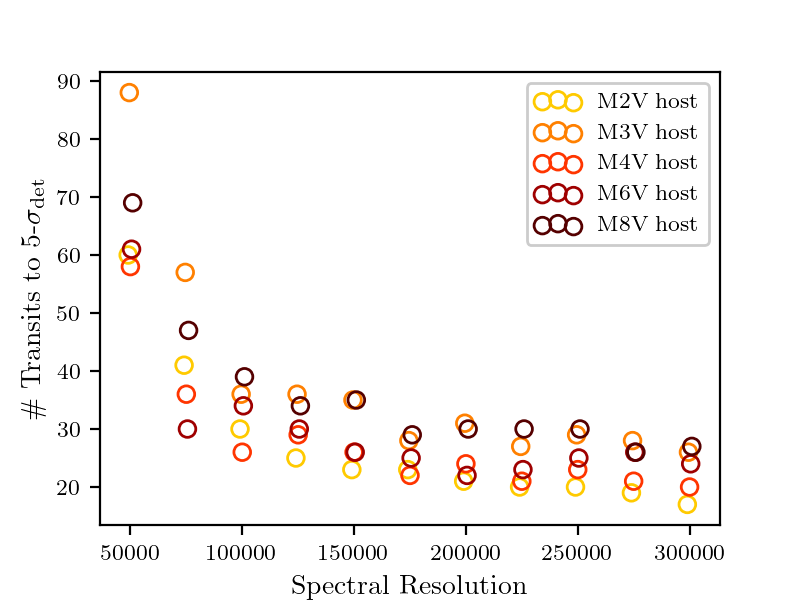}
    \caption{Number of transits required to detect the \ce{O2} 1.27 $\mu$m band as a function of spectral resolution for a PIE atmospheres transiting M dwarf stars 5 pc away from Earth. The benefits of increasing the spectral resolution are marginal for resolutions greater than $R\sim150,000$.}\label{fig:res}
    
\end{figure}

We briefly investigate the effect of varying the spectral resolution of our simulated observations by calculating the number of transits required to detect the \ce{O2} 1.27 $\mu$m band for a PIE planetary atmosphere transiting M2V--M8V host stars 5 pc away from Earth, and present the results in Figure~\ref{fig:res}. We find that decreasing the spectral resolution to $R = 50,000$ roughly doubles the required number of transits, while increasing the resolution to $R \gtrsim 150,000$ only marginally decreases the required number of transits. We note that while average line depth roughly doubles when increasing the spectral resolution from $R=100,000$ to $R=300,000$, this only translates to a ~35\% decrease in the number of transits required for a detection, which is consistent with \citet{Lopez-Morales2019}. The dominant factor controlling the number of transits required for a detection as a function of resolving power is the average SNR of the resolution elements. Assuming the wavelength range remains constant, an increase in spectral resolution requires an increase in the number of pixels per observation, thus the signal is spread over more pixels, and the average SNR of the individual resolution elements decreases. We expect the other molecular bands to roughly follow this trend, and we look forward to completing a more in-depth analysis in our future work. 

\subsubsection{Synthetic spectra}
Our SPECTR pipeline generates synthetic observations by calculating the incoming photon count from the planetary system for both in-transit and out-of-transit observations, and the background photon count from atmospheric molecular emission, airglow, scattered light, zodiacal light, dark current, and read noise sources. We appropriately Doppler shift the planet spectrum over the course of the transit, and multiply each exposure by the corresponding high-resolution telluric transmittance spectrum obtained from SkyCalc.

The stellar photon count rate in counts per unit time is given by:
\begin{equation}\label{eq:cr}
    c_s = T F_s(\lambda) \frac{\lambda}{hc} d\lambda \pi \left( \frac{D}{2} \right)^2
\end{equation}
where $T$ is the total efficiency of the telescope/instrument/detector system multiplied by the wavelength-dependent telluric absorption, $F_s(\lambda)$ is the stellar flux at the top of Earth's atmosphere, $h$ is the Planck constant, $c$ is the speed of light, $d\lambda$ is the spectral element width, and $D$ is the diameter of the telescope. The stellar photon count rate is scaled to simulate an in-transit observation by multiplying by one minus the wavelength-dependent transit depth:
\begin{equation}
    c_\mathrm{in-transit} = c_s \left( 1 - \left( \frac{R_p}{Rs} \right)^2 \right) 
\end{equation}

The atmospheric molecular emission, airglow, scattered light, zodiacal light backgrounds are calculated using the ESO SkyCalc interface, and the dark current and read noise are taken to be typical quoted values for ELT detectors (see Section~\ref{sec:telinstprop}). The thermal contribution to photon count rate is the sum of thermal emission from Earth's atmosphere and the thermal emission of the telescope and instrument, which we set as standard across both telescope/instrument configurations: 273 K for the telescope mirror and 90 K for the instrument/detector, typical standard values for a high-altitude telescope/instrument setup. The noise sources are shown in Figure~\ref{fig:noise}. To minimize the contribution of readout noise to the total noise budget, we fix the exposure time to equal the transit duration for each target. Our planets have relatively short transit durations, and the planetary orbital velocities relative to the host stars are sufficiently small such that any smearing effects on the detector are negligible during the transit. 

Finally, we simulate random Poisson noise in our spectra. We add the signal, background, and dark current to obtain total signal for each resolution element, and total shot noise follows by taking the square root of the total signal. Total noise is then the total shot noise and readout noise added in quadrature. We randomly draw values from Poisson distributions defined by the total noise for each resolution element, and add this to our spectra to simulate noise.

We assume all transit spectra in this study to have a system velocity shift of $22$ km/s relative to Earth (including all sources of velocity shift) to shift the target spectrum away from the telluric transmission lines. This is within the optimal range for reducing blending between telluric and exoplanet \ce{O2} lines at the $0.76\mu m$ absorption feature \citep{Rodler2014}. A non-optimal Doppler shift for the system may increase blending of the planetary and telluric absorption lines by 50\% or more \citep{Rodler2014}.

\subsubsection{A note on host star properties}
Modeling how the properties of a host star affect exoplanet characterization is complex and often unique to each individual system, thus we make simplifications to generalize our study. In particular, we leave out the effects of stellar rotation and star spots. We also do not model the effects of stellar or planetary rotation, but we acknowledge that stellar rotation in particular can make stellar line removal more difficult in transmission via the Rossiter-McLaughlin effect (e.g. \citet{Brogi2016}). 

While some studies have addressed the possibility of star spots interfering with transit studies \citep[][]{Pont2007-wi, Pont2008-gh}, \citet{Brogi2016} showed that the appearance of star spots, at least in the scenario of the hot Jupiter HD 189733 b, was negligible due to the small fraction of spectra which include the spot. However, HD 189733 is a K dwarf, and is thus not as active as the class of star we study in this work. In fact, other studies \citep[e.g.][]{Czesla2009-sd, Desert2011-um, Silva-Valio2011-av, Bruno2016-do} found that both occulted and non-occulted star spots can cause a wavelength-dependent increase or decrease in transit depth, and that in some regions of the spectrum, stellar features can actually be imprinted onto the transmission spectrum \citep{Bruno2020-sv}. This is especially problematic for cool and low-mass stars \citep[e.g.][]{Rackham2017-hw, Rackham2018-yq, Wakeford2018-rr}, which are known to have \ce{H2O} and \ce{CO} in their spectra \citep{Allard1997-ui} that can overlap with the spectral lines of identical molecules in the planetary atmosphere. This molecular overlap effect is likely less problematic for \ce{O2}, \ce{CO2}, \ce{CH4}, which are weak or nonexistent in low-mass stellar spectra \citep{Allard1997-ui}.  While this potential stellar contamination is a further complication to consider when attempting to observe terrestrial exoplanet atmospheres at high spectral resolution, the specific impact is not yet well constrained, and we currently do not include these effects in our simulations.

\subsection{Telluric line removal}\label{sec:telluricremoval}

To remove telluric lines from our spectra, we assume that there exists a ``perfect" corresponding out-of-transit observation at the same airmass for each in-transit observation such that the ratio of the in-transit to out-of-transit spectra leaves only the planetary transmission spectrum and noise. We then construct an outlier mask to flag data in areas of extremely low signal-to-noise (e.g. where the telluric transmittance falls to near zero) by performing a running sigma clip with a width of 100 pixels. All values more than 3$\sigma$ from the median were flagged and not considered in the cross-correlation analysis. To remove low frequency variations associated with the spectral continuum in the template and observed spectra, we apply a high-pass filter with an arbitrarily chosen bin width of $100$ wavelength steps.

However, for real terrestrial exoplanet data, the removal of telluric transmission lines using techniques initially developed for hot Jupiter observations is less likely to be effective, although several other avenues show promise.  Although the radial velocity of the planetary system will Doppler shift the planetary lines away from the telluric lines \citep{Lopez-Morales2019}, the high sensitivity required for these observations suggests additional techniques may be needed to refine telluric subtraction. The current state-of-the-art methods for hot Jupiter telluric transmission line removal employ either a principal component analysis (PCA) algorithm \citep[e.g.][]{Brogi2018-dn}, or a radiative transfer model \citep[e.g.][]{Allart2017-ja} to remove Earth's transmission spectrum. However, for transiting habitable zone terrestrial planets, PCA will be less effective because the planetary velocity shifts during a transit are likely insufficient to effectively isolate the planet spectrum from static telluric lines, resulting in the subtraction of the planet spectrum itself. Therefore, blind analysis techniques like PCA or PCA-like algorithms that do not use existing knowledge of molecular absorption and the Earth's atmospheric properties to subtract tellurics will be less effective for observations of transiting Earth-like planets. 

More promising techniques relevant to terrestrial exoplanet observations include applying radiative transfer tools like Molecfit \citep{Smette2015-jl} or the online TAPAS service \citep{Bertaux2014} to model the Earth’s atmospheric transmission. Molecfit as a telluric line removal tool is demonstrated in \citet{Allart2017-ja}, where it is used to fit and subtract the telluric lines down to the noise level in high-resolution observations of water lines in the atmosphere of HD 189733b. More recently, TAPAS has been used on ESPRESSO data of HD 40307 to fit and remove telluric lines in wavelength regions with significant telluric absorption traditionally excluded by precision radial velocity surveys, improving the precision of the resulting stellar spectrum by up to 25\% \citep{Ivanova2023}. While using a radiative transfer tool to fit the telluric lines of our synthetic data would be more realistic, here we have chosen to present the ideal case with perfect subtraction; however, we look forward to including a more realistic telluric removal process in future iterations of this work.

\subsection{Cross-correlation and detection significance}
To estimate the detectability of the molecular absorption bands, we employ a cross-correlation technique similar to that described in \citet{Brogi2016} for its sensitivity to line locations, line shapes, and relative line depths, and robustness against small perturbations to the radial velocity due to stellar or planetary processes when applied to real data. With the cross-correlation technique, it is not necessary to identify the precise wavelength position of the band before co-adding the flux. We cross-correlate our simulated transmission spectra with a model template of the molecular absorption band based on the techniques tested and used by similar studies \citep[e.g.][]{Snellen2010, Snellen2013, Rodler2014, Brogi2016, Serindag2019, Lopez-Morales2019, Spring2022}, and report detections as the significance at the expected signal location in the resulting cross-correlation functions. 

The cross-correlation technique works by comparing an observed spectrum ($f(n)$) to a range of Doppler-shifted template spectra ($g(n-s)$).  We remove the spectral features of molecules other than the one in question from the template spectrum to reduce contamination, which, left in the spectrum, can lead to inaccurate boosts in the detection significance. For each observation--template comparison, the potential match is quantified with a correlation coefficient, where a more positive coefficient indicates a better match. The Doppler velocity of the template spectrum is allowed to vary over $\pm150$ km/s on an evenly spaced grid of 101 elements. For the grid of velocity shifts, $s$, a cross-correlation function $C(s)$ is calculated from the variance of the observed spectrum ($s_f^2$), the variance of the template (model) spectrum ($s_g^2$), and the cross-covariance $R(s)$. To remove low-frequency variations in the spectral continuum, it is crucial to apply a high-pass filter to the model and observed spectra before the variance, cross-covariance, cross-correlation function are calculated to remove low frequency continuum variations. We adopt the notation of \citet{Brogi2019} and define:

\begin{equation}
    s_f^2 = \frac{1}{N} \sum_n f^2(n)
\end{equation}

\begin{equation}
    s_g^2 = \frac{1}{N} \sum_n g^2(n-s)
\end{equation}

\begin{equation}
    R(s) = \frac{1}{N} \sum_n f(n) g(n - s)
\end{equation}
where n is the bin or pixel number, s is a bin or wavelength shift due to the relative velocity, N is the total number of pixels, $f(n)$ is the synthetic observed spectrum, and $g(n - s)$ is the template spectrum with a wavelength shift of $s$. The cross-correlation function is then defined as
\begin{equation}
    C(s) = \frac{R(s)}{\sqrt{s_f^2 s_g^2}}.
\end{equation}

This results in a cross-correlation function which peaks at a relative velocity of zero if a signal can be detected through the noise. To estimate the detection significance of the resulting cross-correlation function, we follow the $\chi^2$ analysis of \citep{Brogi2016}: We define a non-detection as zero planet signal and zero correlated noise, such that the corresponding distribution of cross-correlation function values is a Gaussian with a mean of zero. Thus, a non-detection is best fit with a flat line with zero offset. To test for a signal, we compare our cross-correlation function to a flat line with zero offset:

\begin{equation}\label{eq:chisq}
    \chi^2_{\mathrm{CCF}} = \sum_{s} \frac{(C(|s| < s_0) - 0)^2}{\sigma^2_{C(|s| > s_0)}}
\end{equation}

In Equation~\ref{eq:chisq}, the sum is only calculated for velocity shifts $|s| < s_0$, where $s_0$ is 5 km/s and chosen to reflect the width of a typical cross-correlation peak in this study, to include only values that could correspond to the planet signal, and exclude any aliasing patterns or spurious matches with other spectral features. Similarly, $\sigma^2_{C(|s| > s_0)}$ is the variance excluding possible planet signal. We convert this $\chi^2_{\mathrm{CCF}}$ value to a p-value using the cumulative distribution function of a $\chi^2$ distribution, and finally convert the p-value to a sigma interval using the inverse survival function of a normal distribution to arrive at a measure of how much our cross-correlation function deviates from a Gaussian distribution, $\sigma_{\mathrm{det}}$.

In this detection significance estimation scheme, a non-detection would be consistent with a detection significance of $\sim1$, which differs from a traditional signal-to-noise ratio non-detection of $0$. To confirm this, we calculated the detection significance of a ``cross-correlation function'' consisting of randomly drawn values from a standard normal distribution. After a million iterations of drawing random CCFs and estimating their detection significances, the median detection significance was $1.04$. Therefore, we expect non-detections in our results to have detection significances consistent with 1.04, and this value is seen at a low number of observed transits or for molecules that are challenging to detect in several of our cases in e.g. Figure~\ref{fig:det_5pc_TMT}. Furthermore, there exists a fundamental upper limit to the detection significance that is unique to each molecular band. Because we estimate detection significance from a cross-correlation function, the largest detection significance possible corresponds to the detection significance of the cross-correlation of the template spectrum and a noiseless observed spectrum. Therefore, instead of obeying a $\sqrt{N_\mathrm{transits}}$ trend indefinitely as a traditional signal-to-noise vs. number of transits observed curve would, we expect our detection curves to ``saturate'' at the detection significance upper limit. Indeed, this trend arises in strong, readily detectable molecular bands in Figure~\ref{fig:det_5pc_TMT}. 

For N transits observed, we integrate N in-transit spectra with random noise and N out-of-transit spectra with random noise, remove the telluric lines via the procedure in Section~\ref{sec:telluricremoval}, and calculate the cross-correlation detection significance. We then repeat this process for 500 iterations. We report the median detection significance with uncertainties corresponding to the standard deviation of the detection significance iterations.

\section{Results}\label{sec:results}
Using our cross-correlation pipeline, we estimate the detectability of molecular bands in terrestrial planet atmospheres transiting M2V through M8V host star types. The atmosphere classes included in our study are self-consistent PIE and ARE atmospheres which contain true biosignatures, and abiotically-generated \ce{O2} via \ce{CO2} photolysis \citep{Harman2018-gq} and ocean-loss/outgassing \citep{Leung2020} atmospheres (see Table~\ref{tab:experiments}). Selected high-resolution spectra of the most detectable molecular absorption bands are presented in Figure~\ref{fig:bands}, and the results for all molecular absorption bands are available as an accompanying data product for this paper. We calculate molecular detectability for these atmospheres for clear- and cloudy-sky scenarios, for GMT, TMT, and E-ELT configurations, and at distances of 5 and 12 pc away from Earth. Although it is unlikely that we find a transiting terrestrial planet at 5 pc, we include simulations of these systems at this distance for direct comparison with previous results that have used 5 pc as a point of reference.

\begin{figure*}
    \centering
    \includegraphics[width=\textwidth]{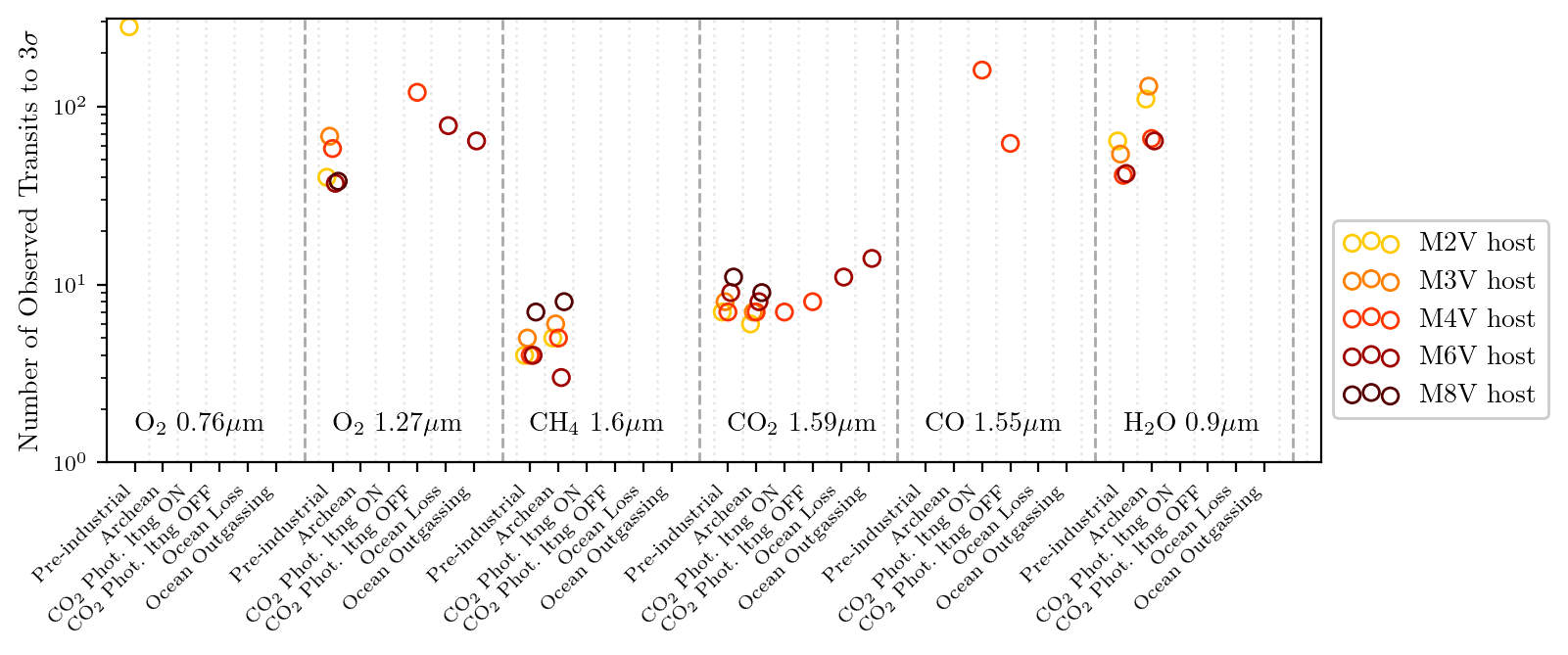}
    \caption{Total observed transits required for a 3-$\sigmadet$ significance detection using an E-ELT sized telescope for selected bands in all atmosphere classes orbiting a range of host stars at a distance of 12 pc from the observer. A marker is missing either where the atmosphere does not contain a molecule or $>300$ observed transits are required to detect the molecular band.}
    \label{fig:dotplot}
\end{figure*}

\begin{figure}
    \centering
    \includegraphics[width=0.45\textwidth]{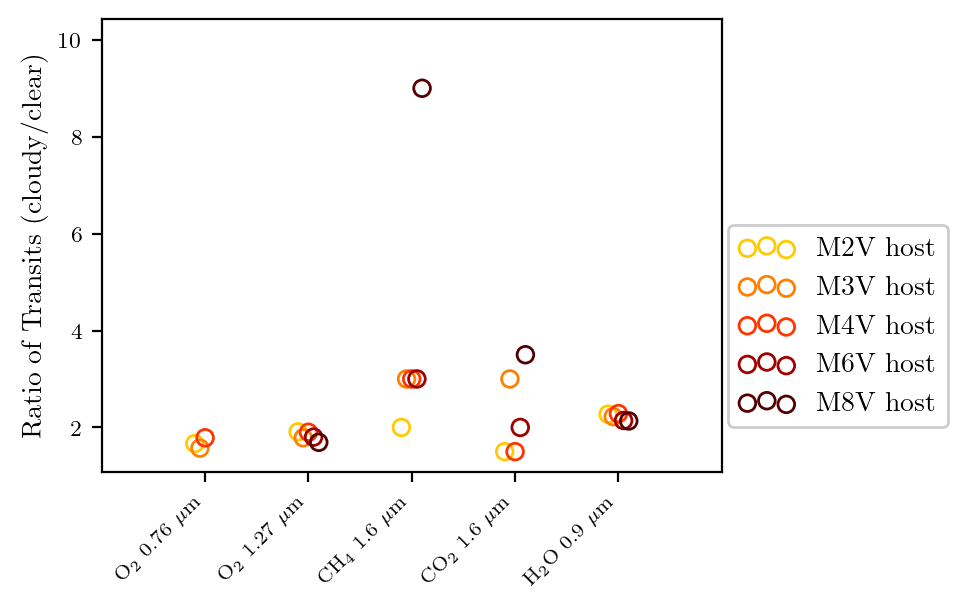}
    \caption{Ratios of the number of transits required to detect the most sensitive molecular bands in our study for cloudy to clear sky pre-industrial Earth-like atmospheres. Missing markers signify that cloudy spectra require $>300$ transits to detect the molecular feature.}
    \label{fig:dotplotclouds}
\end{figure}

\begin{figure*}
    \centering
    \includegraphics[width=0.99\textwidth]{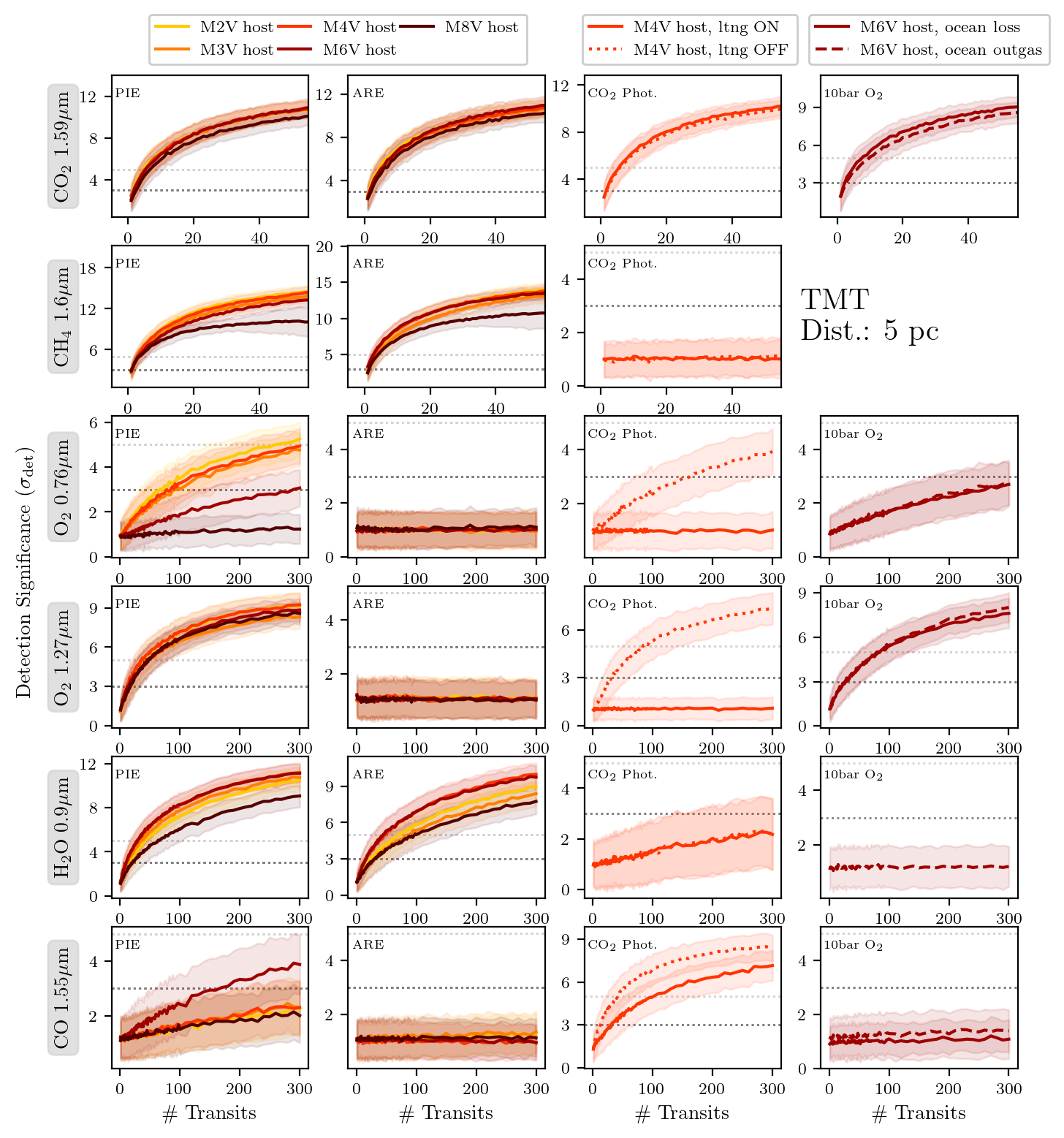}
    \caption{Detection significance as a function of total integration time for the most detectable molecular bands in photochemically self-consistent pre-industrial Earth-like (PIE) atmospheres, Archean Earth-like (ARE) atmospheres, \ce{CO2} photolysis atmospheres (lightning on/off), and 10 bar \ce{O2} worlds (ocean-loss/outgassing) (columns left to right) at distances of 5 pc from Earth observed with a TMT sized telescope. The shaded regions are the 1-$\sigmadet$ uncertainties on detection significance. The dark and light gray dotted lines mark the 3- and 5-$\sigmadet$ thresholds, respectively.}
    \label{fig:det_5pc_TMT}
\end{figure*}

\begin{figure*}
    \centering
    \includegraphics[width=0.99\textwidth]{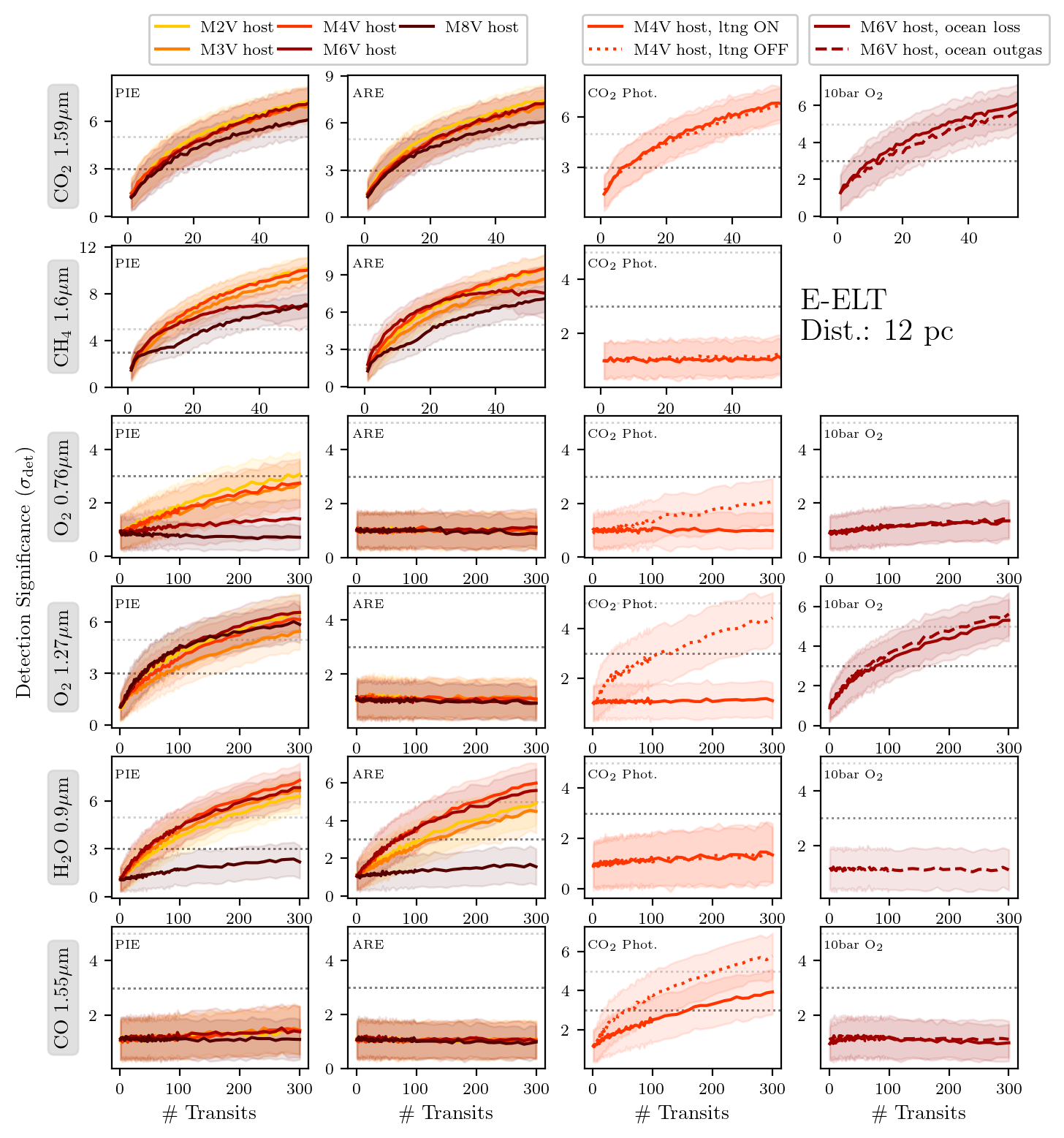}
    \caption{Detection significance as a function of total integration time for the most detectable molecular bands in photochemically self-consistent pre-industrial Earth-like (PIE) atmospheres, Archean Earth-like (ARE) atmospheres, \ce{CO2} photolysis atmospheres (lightning on/off), and 10 bar \ce{O2} worlds (ocean-loss/outgassing) (columns left to right) at distances of 12 pc from Earth observed with an E-ELT sized telescope. The shaded regions are the 1-$\sigmadet$ uncertainties on detection significance. The dark and light gray dotted lines mark the 3- and 5-$\sigma_{\mathrm{det}}$ thresholds, respectively.}
    \label{fig:det_12pc_ELT}
\end{figure*}

\begin{figure*}
    \centering
    \includegraphics[angle=0]{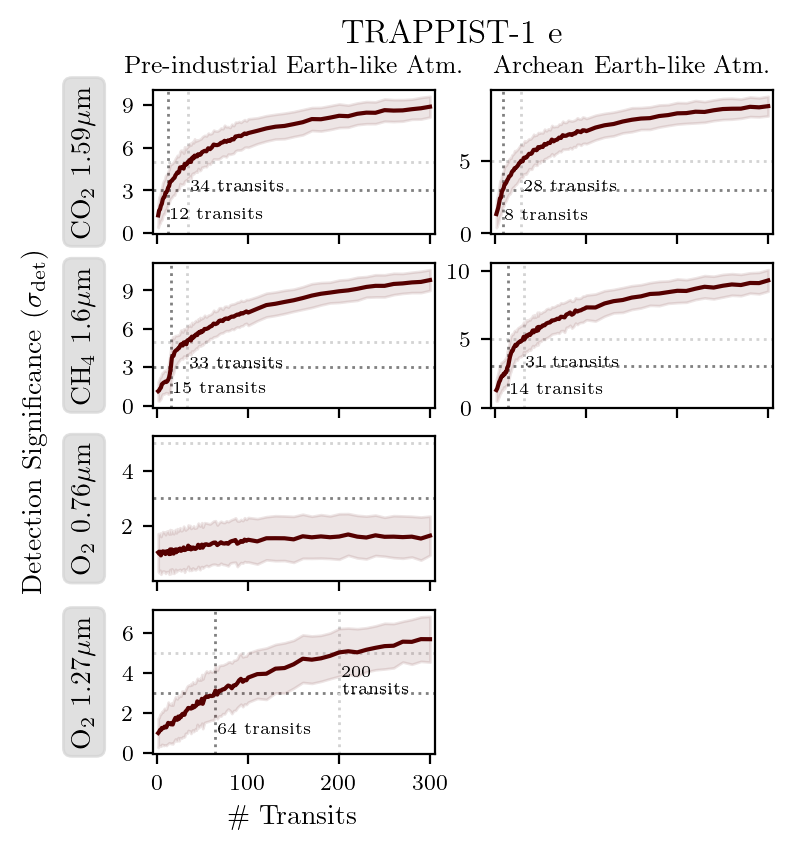}
    \caption{Detection significance as a function of number of observed transits for four molecular bands in our simulated TRAPPIST-1 e atmospheres observed with the E-ELT.  The shaded regions are the 1-$\sigmadet$ uncertainties on detection significance. The horizontal dark and light gray dotted lines mark the 3- and 5-$\sigma_{\mathrm{det}}$ thresholds, respectively. The vertical dotted dark and light gray lines, and corresponding labels to their right, denote the precise integration time required for 3- and 5-$\sigma_{\mathrm{det}}$ detections. Note that the transit duration is 0.99 hr, so integration time required is nearly the same as number of observed transits required.}
    \label{fig:trap1det}
\end{figure*}

A comparison of detectability for the most sensitive bands of specific molecules in a given atmosphere is summarized in Figure~\ref{fig:dotplot}, the effects of clouds for the most sensitive bands are summarized in Figure~\ref{fig:dotplotclouds}. Detailed results on molecular detectability as a function of stellar type, atmosphere type, and number of transits are presented in Figure~\ref{fig:det_5pc_TMT} for targets 5 pc away observed using a TMT configuration and Figure~\ref{fig:det_12pc_ELT} for targets 12 pc away using an E-ELT configuration.  Similarly, we present molecular detectability using an E-ELT configuration for our simulated TRAPPIST-1 e clear-sky atmospheres in Figure~\ref{fig:trap1det}. Each plot shows the detection significance vs. number of observed transits for a molecular absorption band in an atmosphere, orbiting M2V through M8V host stars. Missing lines signify the molecule is not present in the atmosphere, or that the atmosphere is only self-consistent with one host star type, as in the case of the \ce{CO2} photolysis and ocean-loss/outgassing atmospheres (M4V and M6V, respectively). Values for number of transits required for 3- and 5-$\sigmadet$ detections of molecular bands are given in Tables~\ref{tab:results_ELT_pie_clear_12pc} through~\ref{tab:results_falsepos_5pc_TMT}. Although the collecting area of the GMT is much smaller than the E-ELT and TMT, our simulations suggest that some molecular targets, namely \ce{CO2}, will still be feasible to observe with the GMT, and we present these targets in Table~\ref{tab:gmt_results} for systems 12 pc away. The addition of clouds in our atmospheres increases the number of transits required by 2-4x on average (Figure~\ref{fig:dotplotclouds}). 

We apply the same noise properties to an 8.2 m diameter mirror telescope and find that the number of transits required to achieve a similar detection significance increases by at least an order of magnitude.  The distance dependence tracks with the inverse-square law: a target at 10 pc from Earth requires 4 times the number of transits to obtain a 3-$\sigmadet$ significance detection for a target 5 pc away from Earth. 

The most detectable molecules are \ce{CH4} and \ce{CO2}, with their most detectable absorption features residing in the 1.6 $\mu$m region. We find that the $1.27 \mu$m,  $0.9 \mu$m, and $1.55 \mu$m bands are the most detectable \ce{O2}, \ce{H2O}, and \ce{CO} features, respectively. \ce{O3} and \ce{C2H6} were not detectable in any case in this study.

\subsection{Dependence on stellar type}
In nearly all cases, the molecular detectability is dependent on the host star type. 
We identify an interplay between physical and chemical effects of the host star on the detectability of molecular bands in an exoplanet atmosphere. In addition to photochemical effects in the exoplanet atmosphere, the wavelength-dependent brightness and physical size of the stellar disk can influence detectability. 

\subsubsection{Photochemistry}
As discussed in Section~\ref{sec:intro}, the SED of the host star affects both the abundance and distribution of molecules in an exoplanet atmosphere through photochemical processes.  A higher molecular abundance absorbs more photons from the star, creating deeper transits, and leading to stronger detectability. This effect is clearly seen when comparing the \ce{CH4} profiles of self-consistent PIE to true Earth (Figure~\ref{fig:spaghettipie}). Variable UV activity of M dwarf stars allows for the buildup of \ce{CH4}, while \ce{CH4} in the true Earth atmosphere is photochemically destroyed more rapidly \citep{Segura2005-mg}.

\subsubsection{Stellar brightness and size}\label{sec:brightness}
Photochemical effects aside, the wavelength-dependent brightness of the host star SED also impacts the detection significance of individual molecular bands. The incident SED on the planet atmosphere can have wavelength-dependent effects on detectability. For example, molecules that absorb in the region of peak flux in the host star SED have the advantage of more available photons to absorb when compared to an off-peak molecular absorption band, thus producing deeper transits. This effect is illustrated in our results by comparing the $0.76 \mu m$ and $1.27 \mu m$ bands of \ce{O2} in Figure~\ref{fig:det_5pc_TMT}. Since the earlier type M dwarf host SEDs peak closer to the $0.76 \mu m$ band, they have an observational advantage over later-type stars: they require fewer transits to achieve the same detection significance for this band (see Figure~\ref{fig:det_5pc_TMT}). Conversely, later type M dwarf SEDs peak closer to the $1.27 \mu m$ \ce{O2} band, and \ce{O2} is equally or more detectable in this region for late-type hosts. However, in many cases the flux of the host star at the observer due either to intrinsic brightness or proximity (lower panel of Figure~\ref{fig:stellar}) can compete with any wavelength-dependent effects due to the number of photons that are incident on the detector. This effect can be seen in the 1.6 $\mu$m \ce{CH4} band in Figure~\ref{fig:det_12pc_ELT}, where the larger, earlier-type hosts may require fewer transits to achieve a 3- or 5-$\sigmadet$ detection significance, even though the later type host SEDs peak closer to this region. 

Transit depth is also inversely proportional to the radius of the star squared, and so larger stars can reduce the detectability of molecules in the atmospheres of exoplanets transiting them, while smaller stars will enhance the detectability of planetary atmospheric absorption. Deeper transits are easier to distinguish from the surrounding noise, thus are more detectable. This is an important factor working to increase the detectability of molecules in planets transiting late-type hosts.

In summary, detectability is proportional to transit depth and stellar luminosity, which both depend on the size of the host star. Additionally, stellar luminosity depends on the effective temperature of the star. Detectability is also highly variable for different wavelength regions in a transit spectrum due to photochemical effects in the exoplanet atmosphere and telluric effects in the Earth's atmosphere.

\subsection{Favorable targets for the ELTs}

In terms of the fewest number of transits required to detect a molecule, the best overall stellar host targets for characterizing terrestrial-sized exoplanets for habitability and life are M dwarf stars earlier than M6V; however, it is important to note that in terms of least total observing time and opportunities to obtain transmission spectra, later-type M dwarf hosts have the advantage (see Section~\ref{sec:logistics} for a discussion). Since the transit duration of a planet increases with earlier-type hosts, early-type systems require fewer observed transits, while later-type hosts require less overall observation time to achieve the same detection significance. The best targets will depend on the priorities and limitations of the observer. Figure~\ref{fig:dotplot} provides a summary of the number of transits required to detect the most accessible molecular bands for each self-consistent atmosphere transiting M dwarf hosts in our study.

The \ce{CH4} $1.6 \mu m$ and \ce{CO2} $1.59 \mu m$  bands stand out as being the most accessible bands for the ELTs (see Figures~\ref{fig:det_5pc_TMT} and~\ref{fig:det_12pc_ELT}), requiring the fewest number of observed transits (see also Figure~\ref{fig:dotplot}). If a planet atmosphere contains both species, such as our PIE and ARE atmospheres, a $> $5-$\sigmadet$ detection can be achieved for both \ce{CH4} and \ce{CO2} by observing $\sim 20-30$ transits for all stellar host types at a distance of 12 pc for the E-ELT. If abundant \ce{O2} is present in an atmosphere, the band requiring the fewest number of observed transits is the $1.27 \mu m$ band for all stellar host types (see Figure~\ref{fig:det_5pc_TMT}), and can be detected at a detection significance of 5-$\sigmadet$ for PIE atmospheres 12 pc away in $>300$ transits for the GMT, 230 or more transits for the TMT, and 140--180 transits for the E-ELT. The $0.76 \mu m$ A-band may be accessible with significantly more observation time, requiring three or more times the number of observed transits to detect at the same significance as the $1.27 \mu m$ band. We note that this differs from the results of \citep[e.g.][]{Rodler2014, Lopez-Morales2019, Wunderlich2020-fv}, and attribute the differences to their inclusion of red noise sources as 20\% and 50\% of the white noise in the visible and NIR, respectively. Here we choose not to model red noise in this way to simulate an idealistic scenario. \ce{H2O} may also be accessible for planets around any M dwarf host, with the $0.9 \mu$m band as the most detectable feature in all cases, requiring 140--180 transits for a 5-$\sigmadet$ detection for a pre-industrial Earth-like planet transiting an M6V 12 pc away observed with the E-ELT. \ce{CO} may be detectable for mid-type M dwarf hosts in the $1.55 \mu m$ region if sufficient CO is available in the atmosphere (i.e. only for the \ce{CO2} photolysis atmospheres), and is accessible at a 3-$\sigmadet$ level in 160 and 62 transits for scenarios with and without lightning, respectively, for targets 12 pc away observed with the E-ELT. The next most detectable \ce{H2O} and \ce{CO} bands may require an order of magnitude or more observed transits, and may be inaccessible for targets at 12 pc. \ce{O3} and \ce{C2H6} are not detectable with our simulated observational setups.

\subsubsection{TRAPPIST-1 e with the ELTs}\label{sec:trap1results}
To better understand the detectability of molecular bands for a promising known target, we simulate transit transmission observations of PIE and ARE planets transiting TRAPPIST-1 at an orbital geometry analogous to the planet TRAPPIST-1 e. We report the number of transits to 5-$\sigmadet$ for the most detectable molecules in the atmosphere of TRAPPIST-1 e in Table~\ref{tab:resultscompare}, and include detectability estimates for an M4V at 12 pc in the table for comparison. In Figure~\ref{fig:trap1det}, we show that \ce{CO2} and \ce{CH4} are detectable in 34 and 33 observed transits, respectively, at a 5-$\sigmadet$ significance with the E-ELT for a PIE atmosphere, and 28 and 31 transits, respectively, for an ARE atmosphere. Detecting \ce{O2} in a PIE atmosphere will be challenging, requiring 200 observed transits for a 5-$\sigmadet$ significance detection, and 64 transits for a 3-$\sigmadet$ significance detection. \ce{H2O} may be extremely challenging, requiring more than $300$ transits.

\subsection{Molecular targets requiring other observation techniques}
The most difficult molecules to detect with transit transmission spectroscopy using ground-based ELTs are \ce{O3} for all atmospheres, \ce{CO} for PIE and ARE atmospheres, and \ce{C2H6} for ARE atmospheres. The only \ce{O3} bands with wavelength accessibility for the ELTs are the Chappuis band and $3.2 \mu m$ band, and the features do not have sufficient high-resolution structure to use this detection method. CO, which can be an indicator of abiotic generation of \ce{O2}, is most prominent in the \ce{CO2} photolysis atmosphere and may be possible to detect in $<100$ transits in certain scenarios, but in all other cases is not reliably detectable with the ELTs. These molecules may be better observed using space-based telescopes, which can probe other molecular bands, or ground-based reflected light spectroscopy, which can probe deeper into the atmosphere.

\section{Discussion}\label{sec:discussion}
We have identified multiple molecular bands, including \ce{CO2}, \ce{CH4}, \ce{O2}, \ce{H2O}, and \ce{CO}, that could be detectable using the ELTs for terrestrial planet atmospheres transiting a range of M dwarf host stars. The suite of molecules accessible to ground-based observatories could aid in searching for environmental clues that point to life, habitability, and biosignature false-positive mechanisms. Additionally, the factors that affect the detectability of a molecule reach beyond photochemistry and instrument sensitivity, and we note the importance of the host star's physical characteristics, such as size and wavelength-dependent/overall brightness, for selecting ideal systems and molecular bands to observe in the future. We present a recommended observing protocol for discriminating possible terrestrial planet environments below in Section \ref{observingprotocol}

Our results suggest that there are many molecular bands that may be accessible with the ground-based ELTs, and these bands may be able to provide environmental context for an \ce{O2} detection, or point to other outcomes of atmospheric evolution. With our novel cross-correlation pipeline, we estimated the detectability of \ce{O2}, \ce{CH4}, \ce{CO2}, \ce{CO}, \ce{H2O}, \ce{O3}, and \ce{C2H6} in the atmospheres of pre-industrial Earth-like, Archean Earth-like, \ce{CO2} photolysis, and 10 bar ocean-loss/outgassing atmospheres, all photochemically self-consistent with M2 through M8 host stars. We found that \ce{CO2} and \ce{CH4} are excellent targets to search for in suspected habitable Earth-like atmospheres orbiting any type of M dwarf host, taking just a few observed transits for initial discovery. Additionally, \ce{O2} may be detectable for a pre-industrial Earth-like world at both $0.76$ and $1.27 \mu m$, with the $1.27 \mu m$ band having a distinct observational advantage. An atmosphere that builds up abiotic \ce{O2} from \ce{CO2} photolysis may not be able to accumulate detectable levels of \ce{O2} for the more unrealistic case of no lightning in an atmosphere, therefore discrimination from atmospheres influenced by life may not be a problem. However, if \ce{O2} is detected, detection of both \ce{CO2} in 28 transits, and strong \ce{CO} in 84 transits would strongly suggest an abiotic buildup scenario from \ce{CO2} photolysis in the case of a target 12 pc away observed with the E-ELT. 

We compare our \ce{O2} results to previous literature \citep[e.g.][]{Rodler2014, Lopez-Morales2019, Leung2020, Wunderlich2020-fv}, and find that our detectability estimates for planets transiting late-type hosts are in agreement, while our estimates for early-type M dwarf planets differ. We overestimate the 1.27 um detectability, and underestimate the 0.76 um detectability, for early type hosts. We agree on the late-type detectability estimates for both the 0.76 and 1.27 \ce{O2} bands \citep{Lopez-Morales2019}. We hypothesize that this is due to a difference in how we added noise to our simulations: we use ESO SkyCalc to generate realistic wavelength-dependent background noise, while \citep{Rodler2014} and \citep{Lopez-Morales2019} approximate red noise as 20\% and 50\% of the white noise in the visible and NIR, respectively, therefore differences in our detectability estimates are expected.
For planets with significant \ce{O2} buildup due to \ce{H2O} photolysis, \citet{Leung2020} predicts suppression in the 1.27 $\mu$m \ce{O2} band, which we do not see in our results. This is likely explained by a difference in observational methods: we test transmission spectra only, which are more sensitive to the upper layers of the atmosphere and less susceptible to saturate, while \citet{Leung2020} tests reflected light spectra, which probe to the surface and can saturate and suppress the 1.27 $\mu$m \ce{O2} feature.  \citet{Wunderlich2020-fv} investigate the detectability of a similar suite of molecules in Earth-like TRAPPIST-1 e atmospheres with similar abundances for \ce{CH4} and \ce{CO2} to our study using a simple SNR-estimation approach, and they find \ce{CH4} and \ce{CO2} are detectable in 26 and 33 transits, respectively, which are consistent with our estimates of 34 and 33 transits, respectively. However, the \ce{O2} abundance for the \citet{Wunderlich2020-fv} ``wet and alive'' case is 35\%, compared with our 21\% \ce{O2} abundance. Therefore, we cannot directly compare \ce{O2} detectability results, however our results suggest that the $1.27 \mu m$ \ce{O2} band can be detected in $200$ transits, a factor of 4.5 less than \citet{Wunderlich2020-fv}, and we attribute our enhanced \ce{O2} sensitivity to our more rigorous application of the cross-correlation technique, and our ``perfect'' telluric line removal step.

Using the ground-based ELTs, we may have the capacity to search for two biosignature pairs. \ce{O2}, \ce{CH4}, and \ce{CO2} are all detectable within the lifetime of these telescopes at multiple absorption bands, and the \ce{O2}--\ce{CH4} and \ce{CO2}--\ce{CH4} biosignature pairs are potentially accessible. In the most optimistic scenario of a transiting target 12 pc away, both biosignature pairs could be found in as few as 140 observed transits with the E-ELT at 5-$\sigmadet$ significance. The ELTs will require fewer transits to obtain both biosignature pairs than than using JWST, which will require approximately 278 transits for a 5-$\sigma$ detection of the \ce{O2} 1.27 $\mu$m band on TRAPPIST-1 e at 12pc using NIRSpec PRISM  \citep{Meadows2023}. However, the GMT will require $>$300 transits for the \ce{O2} 1.27 $\mu$m, and in this case JWST NIRSpec PRISM may be marginally better. However, the higher-resolution JWST/NIRSpec G140H mode or the NIRISS spectrograph may enhance sensitive to \ce{O2} alone, albeit with a truncated wavelength range, that may be compete with or exceed the capabilities of ground-based observing.

Furthermore, we may be able to discriminate abiotic \ce{O2} buildup scenarios from an Earth-like atmosphere influenced by biology. A detection of \ce{CO} would indicate that the abiotic \ce{O2} is due to \ce{CO2} photolysis by detecting \ce{CO}. A non-detection of \ce{CH4} could also reveal that detected \ce{O2} is abiotic from either \ce{CO2} photolysis or \ce{H2O} photolysis in ocean-loss/outgassing scenarios.

The degree to which it is possible to constrain gas abundances in high-resolution spectra depends on whether the chosen telluric line removal technique also removes continuum information from the observed spectrum \citep{Birkby2018-en}. See Section~\ref{sec:telluricremoval} for a discussion on telluric line removal techniques. One consequence of applying a PCA-like routine is the loss of continuum information, which can introduce degeneracies when comparing the observed spectrum to a grid of atmosphere models to determine molecular abundances. However, given simultaneous observations of multiple gases, it may be possible to infer the relative abundances of the gases since they would depend on the same pressure–temperature profile \citep{De_Kok2013-bv}. On the other hand, knowledge of the Earth’s atmosphere during the observation can inform radiative transfer tools which can model the Earth’s telluric transmission, and remove it from an observed spectrum, preserving the continuum and absolute line depths. Another proposed pathway for preserving the continuum is to combine both low- and high-resolution observations of the same target to break abundance degeneracies \citep{DeKok2014}. The power of multi-resolution observations is demonstrated in \citet[e.g.][]{Pino2018-ur}, where a more sophisticated atmosphere model for HD 189733 b was developed to reconcile the sharp features seen in high-resolution observations with the relatively flat spectrum seen at low-resolution. Abundance determination for terrestrial planets, however, is still an open question to be explored in future work. 

\subsection{Recommended observing protocol} \label{observingprotocol}

To discriminate terrestrial atmosphere types and characterize planetary environments, we recommend prioritizing specific molecular bands to maximize the information gained in as little observation time as possible. Although future designs for high-resolution spectrographs on the ELTs may allow for simultaneous wavelength coverage in the visible \citep[G-CLEF, 0.35-0.95 $\mu$m][]{Szentgyorgyi2014-xj}, NIR \citep[MOHDIS, 0.95-2.4 $\mu$m][]{Mawet2019}, or both \citep[ANDES, 0.4-1.8 $\mu$m][]{Marconi2022-sr}, some of the details of these instruments have not yet been finalized; we therefore treat each molecular band in this study individually in an effort to inform the development of these instruments. In order from highest to lowest priority, we recommend that observers target \ce{CO2}, \ce{CH4}, \ce{O2}, \ce{H2O}, and \ce{CO}. In all cases, there are no detectable \ce{O3} or \ce{C2H6} bands with the ELTs, and \ce{O3} will likely be best detected using the lower-resolution methods of space-based missions. We preface this discussion with the caveat that this study is limited to a handful of planetary atmospheres, and there are many possible outcomes of planetary evolution; detections and non-detections of molecules discussed below should therefore be treated as additional pieces evidence that can increase the probability of certain evolutionary outcomes. Below we synthesize information about each molecule with its relative detectability to justify its placement in our observing protocol priority list.

\subsubsection{\texorpdfstring{\ce{CO2}}{CO2}}
A detection of \ce{CO2} can help rule out larger \ce{H2}-dominated atmospheres by indicating that the atmosphere is primarily the result of planetary outgassing rather than accretion during formation. \ce{CO2} is readily detectable for all of the atmospheres in this work, requiring 20--30 observed transits for a target 12 pc away and 5--10 transits for a target 5 pc away to reach a 5-$\sigmadet$ detection with the E-ELT. 

Additionally, the presence of \ce{CO2} could help constrain planetary climate. \ce{CO2} is a greenhouse gas tightly coupled to geologic processes on our planet. A planet with significant \ce{CO2} may have an active carbonate--silicate cycle \citep{Walker1981-wj, Kasting1993-cn, Kopparapu2013-wm}, which may help buffer the planetary climate through geologic time \citep{Walker1981-wj}. \ce{CO2} could be a useful habitability indicator, however its presence alone is not indicative of an inhabited planet.

\subsubsection{\texorpdfstring{\ce{CH4}}{CH4}}
A \ce{CH4} detection can help discriminate ocean-loss/outgassing and \ce{CO2} photolysis scenarios from Earth-like atmospheres, and provide clues that can point to habitability and life. If a \ce{CO2} detection is made with an accompanying \ce{CH4} detection, the resulting \ce{CH4}/\ce{CO2} disequilibrium pair \citep{Krissansen-Totton2016-bq} may be the most efficient way to search for biosignatures on both PIE and ARE planets, requiring only $\sim 5$ and $\sim 20$ transits in the most optimistic scenarios for transiting Earth-like planets 5 and 12 pc away, respectively. 

\subsubsection{\texorpdfstring{\ce{O2}}{O2}}
Our simulations indicate that the NIR band requires fewer transits to detect \ce{O2} than the A-band for most targets with \ce{O2} in their atmospheres by factors of four or more. A non-detection of \ce{O2} could help rule out PIE and 10 bar \ce{O2} atmospheres, and, if \ce{CO2} is also present, could point to a \ce{CO2} photolysis atmosphere or Archean Earth scenario, although a full suite of gases would be needed to specify the atmosphere type with more certainty. Conversely, a detection of \ce{O2} could provide evidence for a post-ocean-loss, ongoing ocean outgassing, or PIE atmosphere. Furthermore, a simultaneous detection of \ce{CH4} would constitute the \ce{O2}/\ce{CH4} disequilibrium biosignature in a PIE atmosphere. Given that the Archean \ce{CH4}/\ce{CO2} disequilibrium pair is also potentially detectable \citep{Meadows2023}, this means that two potential biosignature disequilibrium pairs, spanning early Earth to modern Earth atmospheres, could be acquired in as few as $\sim 39$ transits in the most optimistic scenario of an Earth-like planet transiting a star 5 pc away. 

\subsubsection{\texorpdfstring{\ce{H2O}}{H2O}}\label{sec:h2o}
A detection of \ce{H2O} can help strengthen the case for a PIE, ARE, or ocean outgassing world, and, conversely, a non-detection can strengthen the case for an ocean-loss or \ce{CO2} photolysis atmosphere. \ce{H2O} detection typically requires a comparable number of transits to the \ce{O2} NIR band to achieve a 3-$\sigmadet$ detection, and a factor of 5--10 or more transits than \ce{CO2} and \ce{CH4}, and for that reason we recommend prioritizing \ce{H2O} after \ce{CO2}, \ce{CH4}, and \ce{O2}. Although the 1-D atmosphere models in our study cannot self-consistently account for 3D effects typically included in 3-D global circulation models (GCMs), the \ce{H2O} profiles are similar enough to first order for reliable detectability estimates \citep{Meadows2023}. 

The 0.9 um band is the most detectable \ce{H2O} band despite the fact that it is weaker than the other bands we tested. Although the \ce{H2O} absorption at 1.1 and 1.3 $\mu$m may be stronger than the 0.9 $\mu$m band in the planet atmosphere, they are also stronger in the telluric transmission spectrum. Compared with the 0.9 $\mu$m region, these longer wavelength regions are dense with \ce{H2O} and overlapping \ce{CH4} lines that saturate to nearly 0\% atmospheric transmission, and effectively block our ability to observe these molecular bands, even at a resolving power of $R = 100,000$. Additionally, transmission spectroscopy is limited in the regions of the atmosphere it can probe; namely, it is more sensitive to upper regions of the atmosphere. To detect strongly absorbing molecules that mainly reside near the surface of an Earth-like planet, like \ce{H2O}, the observer must also overcome this additional challenge, and different techniques such as reflected light observations, which can probe deeper into the atmosphere, may be better suited for detecting \ce{H2O}.

Because an \ce{H2O} detection would be crucial evidence for the presence of liquid water on an exoplanet, we re-ran our analysis combining of all three \ce{H2O} bands in a single spectrum in an attempt to glean any information that may be suppressed in the 1.1 and 1.3 $\mu$m bands. After examining the resulting cross-correlation function, we find that H2O is marginally more detectable with the combined bands, requiring 60 transits for a 3-$\sigmadet$ detection, compared with 64 transits using only the 0.9 $\mu$m \ce{H2O} band. Additionally, the 1.1 and 1.3 $\mu$m bands are out of the nominal GMT G-CLEF wavelength range \citep[0.35-0.95 $\mu$m][]{Szentgyorgyi2014-xj}, and the 0.9 $\mu$m band will be the most suitable target in that case.

\subsubsection{\texorpdfstring{\ce{CO}}{CO}}
A \ce{CO} detection can help identify a world with abiotic \ce{O2} buildup due to \ce{CO2} photolysis, requiring $42$ transits for a planet 5 pc away from Earth. However, \ce{CO} will not be present at significant levels on a planet with life, and should only be prioritized for ruling out an \ce{O2} biosignature false positive scenario. 

\subsubsection{\texorpdfstring{\ce{O3}}{O3} and \texorpdfstring{\ce{C2H6}}{C2H6}}
\ce{O3} and \ce{C2H6} are not accessible using ground-based high-resolution transit transmission spectroscopy.

\begin{deluxetable*}{c|cccc|ccc}\label{tab:resultscompare}
\tablecaption{Number of transits required for 5-$\sigmadet$ detection in modern Earth-like clear-sky atmospheres on TRAPPIST-1 e and around an M4V at 12 pc with the ELTs}
\tablewidth{0pt}
\tablehead{
\colhead{} & \multicolumn{4}{c}{TRAPPIST-1 e} & \multicolumn{3}{c}{M4V at 12 pc} \\
\colhead{} & \colhead{GMT} & \colhead{TMT} & \colhead{E-ELT} & W20 E-ELT &  \colhead{GMT} & \colhead{TMT} & \colhead{E-ELT}
}
\startdata
\ce{CO2} (1.56 $\mu$m) & 240  & 62 & 34 & 33 & 60 & 33 & 25 \\
\ce{CH4} (1.6 $\mu$m)  & $>300$ &  56 & 33 & 26 & $>300$  & 14 & 11 \\
\ce{H2O} (0.9 $\mu$m)  & $>300$  & $>300$ & $>300$ & 1224 & $>300$  & 180 & 130 \\
\ce{O2} (1.27 $\mu$m)  & $>300$  & 300 & 200 & 910 & $>300$ & 240 & 180 
\enddata
\tablecomments{The W20 column presents ELT detectability results from \citet{Wunderlich2020-fv}.}
\end{deluxetable*}

\subsection{Effects of clouds}
We investigate the effects of adding Earth-like clouds to our pre-industrial Earth-like atmospheres. While in most cases we are still able to detect the molecular features in a cloudy spectrum, we find that between two and four times the transits required to detect molecules in the clear sky scenario are required to achieve the same detection significance for cloudy sky scenarios. This is an effect of the clouds suppressing some of the high-resolution features by both raising the spectral continuum in altitude and reducing the relative transit depths \citep{Fauchez2019-sm}. The increase in the number of transits required to detect a molecular feature is a direct consequence of the level of suppression in the spectral features. Of particular note is the \ce{CH4} $1.6 \mu$m band for the PIE atmosphere orbiting the M8V host, which requires nearly 9 times the number of transits to detect this feature than for a clear-sky scenario. 

\subsection{Observing protocol applied to TRAPPIST-1 e}
To better understand how our observing protocol can be used in practice, we apply it to PIE and ARE planets orbiting the M8V host star TRAPPIST-1, simulating transit transmission spectroscopy of TRAPPIST-1 e. In Figure~\ref{fig:trap1det} and Section~\ref{sec:trap1results}, we show that \ce{CO2} and \ce{CH4} are potentially accessible for a TRAPPIST-1 e planet with either a PIE or ARE atmosphere in 12 and 8 transits, respectively, for a 3-$\sigmadet$ detection, and 34 and 28 transits, respectively for a 5-$\sigmadet$ detection. With sufficient observation time, \ce{O2} may be accessible for PIE atmospheres, but would likely require a multi-year observing strategy, even in the most ideal conditions. Similarly, \ce{H2O} would require significant observation time to detect, and may be extremely challenging for the ELTs, even in the case of water vapor transport due to synchronous rotation, as described above in Section~\ref{sec:h2o}. Given the most favorable weather, telescope availability, and seasonal observability, an observer could detect \ce{CO2} and \ce{CH4} in a relatively short timescale, pointing to an Earth-like atmosphere and revealing a biosignature pair. 

\subsubsection{Complementarities to space-based missions}
The ELTs may be powerful tools for terrestrial exoplanet characterization, and complementary to current and future space-based missions. JWST can be used to characterize transiting rocky exoplanets for the biosignature gases \ce{CO2} and \ce{CH4}, but \ce{O2} will be more challenging \citep{Lustig-Yaeger2019-bk, Pidhorodetska2020-gv, Krissansen-Totton2018-wk, Wunderlich2019}.

We compare our ELT detectability results to similar studies for an Earth-like TRAPPIST-1 e observed with JWST, and find that JWST detectability results vary depending on whether the atmospheric composition used was photochemically and climatically self-consistent with the parent star. While it is not straightforward to compare our results to current estimates for JWST detectability due to the different specified atmospheres for these studies (e.g. \ce{CO2} abundances that span 400 ppm to 10\%) we can look at the broad range of estimated detectability anticipated over a broad range of atmospheric gas abundances. Studies with photochemically self-consistent TRAPPIST-1 e Earth-like atmospheres \citep[e.g.][]{Lustig-Yaeger2019-bk, Krissansen-Totton2018-wk, Pidhorodetska2020-gv, Gialluca2021-cq, Wunderlich2019, Wunderlich2020-fv} use a range of \ce{CO2} abundances, and estimate that in atmospheres with 400 ppm of \ce{CO2}, it is detectable in $\sim 10$ transits \citep{Lustig-Yaeger2019-bk}, while a more analogous atmosphere to our study with 10\% \ce{CO2} may be detectable in 5 transits \citep{Wunderlich2020-fv}, roughly 7 times fewer transits than for the ELTs. One consequence of studies that consider non-photochemically self-consistent atmospheres is that even biological amounts of \ce{CH4} may not reach high enough abundances to be readily detectable, requiring $\sim 55$ or more transits to reach a detection \citep{Fauchez2019-sm, Tremblay2020}, compared with 34 transits for our self-consistent E-ELT results. Therefore, while there are relatively few studies which provide direct comparisons between JWST and ELT capabilities, it is apparent that the ELTs will be powerful tools for followup or simultaneous observations to confirm detections of \ce{CO2} or \ce{CH4}, or to dedicate significant ELT resources to search for molecules like \ce{O2} that may be out of reach for JWST. 

Ground-based high-resolution spectroscopy is additionally more sensitive to line cores higher in the atmosphere, and may be used to probe above the cloud deck \citep{Gandhi2020-ln}, while low-to-mid resolution space-based instruments will be limited to relatively few spectral features for cloudy atmospheres \citep{Fauchez2019-sm}. Compared to low-resolution observations, the cross-correlation technique is uniquely sensitive to both the relative depths and spacing of individual absorption lines as well as the high-resolution structure of the lines in the full molecular band, potentially enhancing our ability to break degeneracies due to overlapping absorption bands in low-resolution observations. Without a coronagraph designed for terrestrial exoplanet characterization, JWST observations will be limited to transmission spectroscopy only. Conversely, terrestrial exoplanets are being considered in the design for the ELT coronagraphs, and the ELTs will have access to a larger number of targets with the addition of reflected light spectroscopy as a capability, a topic we leave for future work. 

Looking ahead to the next generation of space-based instruments, the Astro2020 Decadal Survey recommended a 6 m class space mission capable of directly imaging terrestrial-sized exoplanets as a top science priority (Decadal Survey on Astronomy and Astrophysics 2021). Obtaining a direct spectrum of a terrestrial exoplanet will require a coronagraph designed for high-contrast imaging of habitable zone exoplanets, and the technology requirements set by this goal can be implemented and tested with the current designs for ELT high-contrast imaging capabilities in the near term. The lessons learned from the ground will be crucial in the design process for a space-based direct imaging mission, and may help inform future observing strategies. 

\subsection{Observing logistics and caveats}\label{sec:logistics}
Although seasonal observability is a constraint for any observatory, space-based telescopes are unencumbered by factors that limit ground-based observing programs, including variation in the weather and the day-to-night cycle, and may allow more opportunities to observe a transiting planet.  However, \citet{Lovis2017} estimate that despite the restrictions of ground-based observing, it may be possible to detect \ce{O2} in the atmosphere of the non-transiting Proxima Centauri b at 3.6$\sigma$ in 60 nights spread over three years, assuming observations are obtained twice per orbit at approaching and receding quadrature. For comparison with this non-transiting planet at 1.3 pc from Earth using an 8.2 m telescope, we have calculated the feasibility of detecting \ce{O2} from the ground for the transiting HZ planet, TRAPPIST-1 e at 12 pc with an ELT, where 64 transits are needed for a 3-$\sigmadet$ detection. TRAPPIST-1 e undergoes transit once every 6.1 days, corresponding to 60 transits per year; however, using the Exoplanet Archive transit observability prediction tool with a maximum airmass of 3 predicts that only $\sim10$ of these (17\%) are observable from the ground \citep{Akeson2013}. If 35\% of a semester's observing time is also lost to weather\footnote{this statistic was derived from Gemini-S (https://www.gemini.edu/observing/science-operations-statistics)}, this means that only $\sim6$ transits can be observed per year. Thus a 3-$\sigmadet$ detection of \ce{O2} in the PIE scenario of TRAPPIST-1 e will take $\sim 10$ years, and $\sim 33$ years for a 5-$\sigmadet$ detection. If this system was instead at a distance of 5 pc, these estimates would improve to $\sim 2$ years and $\sim 6.5$ years for 3- and 5-$\sigmadet$ detections, respectively.  This real-time estimate potentially increases for habitable zone planets transiting early type hosts, where orbital periods can reach 30 days or more, and the resulting increased transit durations for early-type M dwarf planets may not fully compensate for the decreased number of opportunities to observe a transit. Consequently, while likely possible from the ground, high-resolution spectral exploration of transiting HZ terrestrial exoplanet atmospheres for the known planets at 12 pc will remain challenging, and will require a significant investment in observing time over several years, and possibly decades. In comparison, the closer, non-transiting planets may provide faster access to adequate S/N for HZ exoplanet atmospheres, and we will explore these in a subsequent paper.

\section{Conclusion}\label{sec:conclusion}

The ELTs will be capable of detecting other molecules that can help place detected \ce{O2} in its environmental context, including \ce{CO2}, \ce{CH4}, \ce{H2O}, and \ce{CO}. For the terrestrial planets considered here, \ce{CO2} and \ce{CH4} stand out as the best molecular targets for the ELTs. In particular, a \ce{CO2} detection can increase the likelihood that a planet is terrestrial, and searching for \ce{CO2} may also reveal biological levels of \ce{CH4} (if present), constituting a biosignature pair. A non-detection of \ce{CH4} may point to an abiotic environment. Additionally, detecting Earth-like levels of \ce{O2} in an atmosphere will be challenging, but not impossible for the known transiting targets at 12 pc, and more feasible for any transiting M dwarf planets found at closer distances. However, since \ce{O2} is likely not accessible with JWST, ground-based searches for \ce{O2} provide the best possible means of detecting this important molecule, which can potentially identify an Earth-like planet with oxygenic photosynthesis, and reveal a second biosignature pair (\ce{O2}/\ce{CH4}) if \ce{CH4} can also be detected.  This pair will help to rule out a scenario where abiotic \ce{O2} has built up to detectable levels. A simultaneous non-detection of \ce{H2O} or \ce{CH4} may also help discriminate an abiotic \ce{O2} scenario from a planet with biogenic \ce{O2}. Furthermore, a detection of \ce{CO} may help discriminate abiotic \ce{O2} buildup via \ce{CO2} photolysis from \ce{O2} buildup via ocean loss or outgassing. 

The ELTs have the potential to be used to reveal two biosignature pairs, and detect the molecules necessary to discriminate biological and false positive origins for \ce{O2}.  They will be even more powerful when techniques for telluric line subtraction are improved, and if transiting targets are found closer than the existing ones at 12 pc.  In addition to testing and demonstrating technology that will be crucial to the success of future space-based direct imaging missions, their versatility and relatively near-term availability is complementary to current space-based missions, and they will provide opportunities to observe planetary targets and spectral features that are otherwise inaccessible in the near term.

\section{Acknowledgments}
The authors thank our anonymous reviewers for their comments and suggestions that substantially improved the clarity and robustness of the paper. We would also like to thank Motohide Tamura (University of Tokyo) for reading an early version of this paper and providing valuable comments that helped to strengthen the paper.  M.C. also thanks Michael Line (Arizona State University) and Matteo Brogi (University of Warwick) for helpful discussions on the capabilities and limitations of high-resolution spectroscopy. This work was performed by the Virtual Planetary Laboratory Team, a member of the NASA Nexus for Exoplanet System Science (NExSS), funded via NASA Astrobiology Program Grant No. 80NSSC18K0829, and this work benefited from our participation in the NExSS research coordination network. This work was also partly supported by the Astrobiology Center, Japan. The simulations in this work were facilitated though the use of advanced computational, storage, and networking infrastructure provided by the Hyak supercomputer system at the University of Washington.  

\appendix
\section{Appendix}
Below we include detailed tables for the detectability estimates of each atmosphere type in our study for the configurations of observing targets 5 pc away with the TMT, and targets 12 pc away with the E-ELT. Tables~\ref{tab:results_ELT_pie_clear_12pc} through~\ref{tab:results_falsepos_5pc_TMT} show exact values for the number of transits required to detect each molecule at detection significances of 3-$\sigmadet$ and 5-$\sigmadet$. Table~\ref{tab:gmt_results} shows the targets at 12 pc away that will be accessible to the GMT. Ellipses are given for molecular bands not detectable with the ELTs.

% pie tables

\begin{deluxetable*}{cc|ccccc}[h]\label{tab:results_ELT_pie_clear_12pc}
\tablewidth{0pt}
\tablecaption{Detectability results for clear sky pre-industrial Earths at 12 pc with an E-ELT sized telescope. Missing values indicate that the molecular band was not detectable at the missing significance level in less than 300 transits.}\tablehead{
\colhead{Molecule} & \colhead{Band} & \colhead{M2V} & \colhead{M3V} & \colhead{M4V} & \colhead{M6V} & \colhead{M8V} \\ 
\colhead{} & \colhead{} & \multicolumn{5}{c}{\# transits to 3-$\sigmadet$ (5-$\sigmadet$) detection}
}
\startdata
\multirow{3}{*}{\ce{O2}} & 0.69$\mu m$ & \nodata & \nodata & \nodata & \nodata & \nodata \\ 
 & 0.76$\mu m$ & 280 & \nodata & \nodata & \nodata & \nodata \\ 
 & 1.27$\mu m$ & 40(160) & 68(240) & 58(180) & 37(140) & 38(150) \\ 
\hline\multirow{4}{*}{\ce{CH4}} & 0.89$\mu m$ & 210 & 250 & 220 & \nodata & \nodata \\ 
 & 1.1$\mu m$ & \nodata & \nodata & \nodata & \nodata & \nodata \\ 
 & 1.3$\mu m$ & 48(140) & 68(210) & 64(200) & 80(220) & 120 \\ 
 & 1.6$\mu m$ & 4(11) & 5(12) & 4(11) & 4(12) & 7(25) \\ 
\hline\multirow{2}{*}{\ce{CO2}} & 1.59$\mu m$ & 7(20) & 8(24) & 7(25) & 9(25) & 11(32) \\ 
 & 2.0$\mu m$ & 12(35) & 16(52) & 20(62) & \nodata & \nodata \\ 
\hline\multirow{3}{*}{\ce{H2O}} & 0.9$\mu m$ & 64(180) & 54(160) & 41(130) & 42(140) & \nodata \\ 
 & 1.1$\mu m$ & \nodata & \nodata & \nodata & \nodata & \nodata \\ 
 & 1.3$\mu m$ & \nodata & \nodata & \nodata & \nodata & \nodata \\ 
\hline\multirow{2}{*}{\ce{CO}} & 1.55$\mu m$ & \nodata & \nodata & \nodata & \nodata & \nodata \\ 
 & 2.3$\mu m$ & \nodata & \nodata & \nodata & \nodata & \nodata \\ 
\hline\multirow{3}{*}{\ce{O3}} & 0.63$\mu m$ & \nodata & \nodata & \nodata & \nodata & \nodata \\ 
 & 0.65$\mu m$ & \nodata & \nodata & \nodata & \nodata & \nodata \\ 
 & 3.2$\mu m$ & \nodata & \nodata & \nodata & \nodata & \nodata \\ 
\hline\enddata 
\end{deluxetable*}

\begin{deluxetable*}{cc|ccccc}[h]\label{tab:results_TMT_pie_clear_5pc}
\tablewidth{0pt}
\tablecaption{Detectability results for clear sky pre-industrial Earths at 5 pc with a TMT sized telescope. Missing values indicate that the molecular band was not detectable at the missing significance level in less than 300 transits.}\tablehead{
\colhead{Molecule} & \colhead{Band} & \colhead{M2V} & \colhead{M3V} & \colhead{M4V} & \colhead{M6V} & \colhead{M8V} \\ 
\colhead{} & \colhead{} & \multicolumn{5}{c}{\# transits to 3-$\sigmadet$ (5-$\sigmadet$) detection}
}
\startdata
\multirow{3}{*}{\ce{O2}} & 0.69$\mu m$ & 180 & 280 & 230 & \nodata & \nodata \\ 
 & 0.76$\mu m$ & 68(260) & 88 & 78 & 290 & \nodata \\ 
 & 1.27$\mu m$ & 15(39) & 17(47) & 13(37) & 13(48) & 18(45) \\ 
\hline\multirow{4}{*}{\ce{CH4}} & 0.89$\mu m$ & 49(150) & 62(180) & 50(160) & 96 & \nodata \\ 
 & 1.1$\mu m$ & \nodata & \nodata & \nodata & \nodata & \nodata \\ 
 & 1.3$\mu m$ & 12(34) & 19(52) & 17(48) & 21(68) & 37(110) \\ 
 & 1.6$\mu m$ & 1(3) & 2(4) & 2(3) & 1(4) & 2(4) \\ 
\hline\multirow{2}{*}{\ce{CO2}} & 1.59$\mu m$ & 2(5) & 2(6) & 2(6) & 2(6) & 3(7) \\ 
 & 2.0$\mu m$ & 3(7) & 3(9) & 3(9) & \nodata & \nodata \\ 
\hline\multirow{3}{*}{\ce{H2O}} & 0.9$\mu m$ & 15(42) & 13(36) & 11(31) & 11(29) & 21(64) \\ 
 & 1.1$\mu m$ & \nodata & \nodata & \nodata & \nodata & \nodata \\ 
 & 1.3$\mu m$ & 120 & 130 & 140 & 58(180) & 140 \\ 
\hline\multirow{2}{*}{\ce{CO}} & 1.55$\mu m$ & \nodata & \nodata & \nodata & 170 & \nodata \\ 
 & 2.3$\mu m$ & \nodata & \nodata & \nodata & \nodata & \nodata \\ 
\hline\multirow{3}{*}{\ce{O3}} & 0.63$\mu m$ & \nodata & \nodata & \nodata & \nodata & \nodata \\ 
 & 0.65$\mu m$ & \nodata & \nodata & \nodata & \nodata & \nodata \\ 
 & 3.2$\mu m$ & \nodata & \nodata & \nodata & \nodata & \nodata \\ 
\hline\enddata 
\end{deluxetable*}

% arch tables

\begin{deluxetable*}{cc|ccccc}[h]\label{tab:results_ELT_arch_clear_12pc}
\tablewidth{0pt}
\tablecaption{Detectability results for clear sky Archean Earths at 12 pc with an E-ELT sized telescope. Missing values indicate that the molecular band was not detectable at the missing significance level in less than 300 transits.}\tablehead{
\colhead{Molecule} & \colhead{Band} & \colhead{M2V} & \colhead{M3V} & \colhead{M4V} & \colhead{M6V} & \colhead{M8V} \\ 
\colhead{} & \colhead{} & \multicolumn{5}{c}{\# transits to 3-$\sigmadet$ (5-$\sigmadet$) detection}
}
\startdata
\multirow{3}{*}{\ce{O2}} & 0.69$\mu m$ & \nodata & \nodata & \nodata & \nodata & \nodata \\ 
 & 0.76$\mu m$ & \nodata & \nodata & \nodata & \nodata & \nodata \\ 
 & 1.27$\mu m$ & \nodata & \nodata & \nodata & \nodata & \nodata \\ 
\hline\multirow{4}{*}{\ce{CH4}} & 0.89$\mu m$ & \nodata & \nodata & 300 & \nodata & \nodata \\ 
 & 1.1$\mu m$ & \nodata & \nodata & \nodata & \nodata & \nodata \\ 
 & 1.3$\mu m$ & 58(170) & 82(240) & 70(190) & 68(190) & 150 \\ 
 & 1.6$\mu m$ & 5(13) & 6(16) & 5(12) & 3(10) & 8(23) \\ 
\hline\multirow{2}{*}{\ce{CO2}} & 1.59$\mu m$ & 6(18) & 7(23) & 7(21) & 8(22) & 9(29) \\ 
 & 2.0$\mu m$ & 12(34) & 16(50) & 18(62) & \nodata & \nodata \\ 
\hline\multirow{3}{*}{\ce{H2O}} & 0.9$\mu m$ & \nodata & \nodata & 66(210) & 64(240) & \nodata \\ 
 & 1.1$\mu m$ & \nodata & \nodata & \nodata & \nodata & \nodata \\ 
 & 1.3$\mu m$ & \nodata & \nodata & \nodata & \nodata & \nodata \\ 
\hline\multirow{2}{*}{\ce{CO}} & 1.55$\mu m$ & \nodata & \nodata & \nodata & \nodata & \nodata \\ 
 & 2.3$\mu m$ & \nodata & \nodata & \nodata & \nodata & 41 \\ 
\hline\multirow{3}{*}{\ce{O3}} & 0.63$\mu m$ & \nodata & \nodata & \nodata & \nodata & \nodata \\ 
 & 0.65$\mu m$ & \nodata & \nodata & \nodata & \nodata & \nodata \\ 
 & 3.2$\mu m$ & \nodata & \nodata & \nodata & \nodata & \nodata \\ 
\hline\multirow{1}{*}{\ce{C2H6}} & 3.33$\mu m$ & \nodata & \nodata & \nodata & \nodata & \nodata \\ 
\hline\enddata 
\end{deluxetable*}

\begin{deluxetable*}{cc|ccccc}[h]\label{tab:results_TMT_arch_clear_5pc}
\tablewidth{0pt}
\tablecaption{Detectability results for clear sky Archean Earths at 5 pc with a TMT sized telescope. Missing values indicate that the molecular band was not detectable at the missing significance level in less than 300 transits.}\tablehead{
\colhead{Molecule} & \colhead{Band} & \colhead{M2V} & \colhead{M3V} & \colhead{M4V} & \colhead{M6V} & \colhead{M8V} \\ 
\colhead{} & \colhead{} & \multicolumn{5}{c}{\# transits to 3-$\sigmadet$ (5-$\sigmadet$) detection}
}
\startdata
\multirow{3}{*}{\ce{O2}} & 0.69$\mu m$ & \nodata & \nodata & \nodata & \nodata & \nodata \\ 
 & 0.76$\mu m$ & \nodata & \nodata & \nodata & \nodata & \nodata \\ 
 & 1.27$\mu m$ & \nodata & \nodata & \nodata & \nodata & \nodata \\ 
\hline\multirow{4}{*}{\ce{CH4}} & 0.89$\mu m$ & 78(220) & 130 & 74(220) & 160 & \nodata \\ 
 & 1.1$\mu m$ & \nodata & \nodata & \nodata & \nodata & \nodata \\ 
 & 1.3$\mu m$ & 16(44) & 23(60) & 18(45) & 21(62) & 46(140) \\ 
 & 1.6$\mu m$ & 2(4) & 2(5) & 2(4) & 1(3) & 2(5) \\ 
\hline\multirow{2}{*}{\ce{CO2}} & 1.59$\mu m$ & 2(5) & 2(6) & 2(5) & 2(5) & 3(6) \\ 
 & 2.0$\mu m$ & 3(7) & 3(8) & 3(9) & 36 & \nodata \\ 
\hline\multirow{3}{*}{\ce{H2O}} & 0.9$\mu m$ & 24(72) & 30(90) & 17(48) & 17(45) & 31(98) \\ 
 & 1.1$\mu m$ & \nodata & \nodata & \nodata & \nodata & \nodata \\ 
 & 1.3$\mu m$ & 260 & 240 & 220 & 160 & \nodata \\ 
\hline\multirow{2}{*}{\ce{CO}} & 1.55$\mu m$ & \nodata & \nodata & \nodata & \nodata & \nodata \\ 
 & 2.3$\mu m$ & \nodata & \nodata & \nodata & \nodata & \nodata \\ 
\hline\multirow{3}{*}{\ce{O3}} & 0.63$\mu m$ & \nodata & \nodata & \nodata & \nodata & \nodata \\ 
 & 0.65$\mu m$ & \nodata & \nodata & \nodata & \nodata & \nodata \\ 
 & 3.2$\mu m$ & \nodata & \nodata & \nodata & \nodata & \nodata \\ 
\hline\multirow{1}{*}{\ce{C2H6}} & 3.33$\mu m$ & \nodata & \nodata & \nodata & \nodata & \nodata \\ 
\hline\enddata 
\end{deluxetable*}

\begin{deluxetable*}{cc|cc|cc}\label{tab:results_falsepos_12pc_ELT}
\tablewidth{0pt}
\tablecaption{Detectability results for clear sky \ce{O2} false positive cases with an E-ELT sized telescope at 12 pc. Missing values indicate that the molecular band was not detectable at the missing significance level in less than 300 transits.}\tablehead{
\colhead{\multirow{2}{*}{Molecule} } & \colhead{\multirow{2}{*}{Band} } & \colhead{\multirow{2}{*}{\vtop{\hbox{\strut M4V host}\hbox{\strut \ce{CO2} Phot.}\hbox{\strut ltng. ON}}}}  & \colhead{\multirow{2}{*}{\vtop{\hbox{\strut M4V host}\hbox{\strut \ce{CO2} Phot.}\hbox{\strut ltng. OFF}}}} & \colhead{\multirow{2}{*}{\vtop{\hbox{\strut M6V host}\hbox{\strut 10 bar \ce{O2}}\hbox{\strut ocean outgas}}}} & \colhead{\multirow{2}{*}{\vtop{\hbox{\strut M6V host}\hbox{\strut 10 bar \ce{O2}}\hbox{\strut ocean-loss}}}} \\ 
\colhead{} & \colhead{} & \colhead{} & \colhead{} & \colhead{} & \colhead{} \\
\colhead{} & \colhead{} & \multicolumn{4}{c}{\# transits to 3-$\sigmadet$ (5-$\sigmadet$) detection}
}
\startdata
\multirow{3}{*}{\ce{O2}}        & 0.69$\mu m$    & \nodata      & \nodata       & \nodata & \nodata \\ 
                                & 0.76$\mu m$    & \nodata      & \nodata       & \nodata & \nodata \\ 
                                &  1.27$\mu m$   & \nodata      & 120           & 64(240) & 78(270) \\ 
\hline\multirow{4}{*}{\ce{CH4}} & 0.89$\mu m$    & \nodata      & \nodata       & \nodata & \nodata \\ 
                                & 1.1$\mu m$     & \nodata      & \nodata       & \nodata & \nodata \\ 
                                & 1.3$\mu m$     & \nodata      & \nodata       & \nodata & \nodata \\ 
                                & 1.6$\mu m$     & \nodata      & \nodata       & \nodata & \nodata \\ 
\hline\multirow{2}{*}{\ce{CO2}} & 1.59$\mu m$    & 7(25)        & 8(28)         & 14(40)   & 11(34) \\ 
                                & 2.0$\mu m$     & 21(86)       & 21(84)        & \nodata   & \nodata \\ 
\hline\multirow{2}{*}{\ce{CO}}  & 1.55$\mu m$    & 160          & 62(210)       & \nodata & \nodata \\ 
                                & 2.3$\mu m$     & \nodata      & \nodata       & \nodata & \nodata \\ 
\hline\multirow{3}{*}{\ce{H2O}} & 0.9$\mu m$     & \nodata      & \nodata       & \nodata & \nodata \\ 
                                & 1.1$\mu m$     & \nodata      & \nodata       & \nodata & \nodata \\ 
                                & 1.3$\mu m$     & \nodata      & 210           & \nodata & \nodata \\
\hline\enddata 
\end{deluxetable*}

% false pos table pie
\begin{deluxetable*}{cc|cc|cc}\label{tab:results_falsepos_5pc_TMT}
\tablewidth{0pt}
\tablecaption{Detectability results for clear sky \ce{O2} false positive cases with a TMT sized telescope at 5 pc. Missing values indicate that the molecular band was not detectable at the missing significance level in less than 300 transits.}\tablehead{
\colhead{\multirow{2}{*}{Molecule} } & \colhead{\multirow{2}{*}{Band} } & \colhead{\multirow{2}{*}{\vtop{\hbox{\strut M4V host}\hbox{\strut \ce{CO2} Phot.}\hbox{\strut ltng. ON}}}}  & \colhead{\multirow{2}{*}{\vtop{\hbox{\strut M4V host}\hbox{\strut \ce{CO2} Phot.}\hbox{\strut ltng. OFF}}}} & \colhead{\multirow{2}{*}{\vtop{\hbox{\strut M6V host}\hbox{\strut 10 bar \ce{O2}}\hbox{\strut ocean outgas}}}} & \colhead{\multirow{2}{*}{\vtop{\hbox{\strut M6V host}\hbox{\strut 10 bar \ce{O2}}\hbox{\strut ocean-loss}}}} \\ 
\colhead{} & \colhead{} & \colhead{} & \colhead{} & \colhead{} & \colhead{} \\
\colhead{} & \colhead{} & \multicolumn{4}{c}{\# transits to 3-$\sigmadet$ (5-$\sigmadet$) detection}
}
\startdata
\multirow{3}{*}{\ce{O2}}        & 0.69$\mu m$    & \nodata  & \nodata       & \nodata & \nodata \\ 
                                & 0.76$\mu m$    & \nodata  & 170           & \nodata & \nodata \\ 
                                &  1.27$\mu m$   & \nodata  & 31(84)        & 29(80) & 27(86) \\ 
\hline\multirow{4}{*}{\ce{CH4}} & 0.89$\mu m$    & \nodata  & \nodata       & \nodata & \nodata \\ 
                                & 1.1$\mu m$     & \nodata  & \nodata       & \nodata & \nodata \\ 
                                & 1.3$\mu m$     & \nodata  & \nodata       & \nodata & \nodata \\ 
                                & 1.6$\mu m$     & \nodata  & \nodata       & \nodata & \nodata \\ 
\hline\multirow{2}{*}{\ce{CO2}} & 1.59$\mu m$    & 2(6)     & 2(6)          & 3(10)   & 3(8) \\ 
                                & 2.0$\mu m$     & 3(10)    & 3(10)         & \nodata   & \nodata \\ 
\hline\multirow{2}{*}{\ce{CO}}  & 1.55$\mu m$    & 32(110)  & 14(42)        & \nodata & \nodata \\ 
                                & 2.3$\mu m$     & \nodata  & \nodata       & \nodata & \nodata \\ 
\hline\multirow{3}{*}{\ce{H2O}} & 0.9$\mu m$     & \nodata  & \nodata       & \nodata & \nodata \\ 
                                & 1.1$\mu m$     & \nodata  & \nodata       & \nodata & \nodata \\ 
                                & 1.3$\mu m$     & 78(270)  & 52(170)       & \nodata & \nodata \\
\hline\enddata 
\end{deluxetable*}

% GMT 12 pc results
\begin{deluxetable*}{cccccc}\label{tab:gmt_results}
\tablewidth{0pt}
\tablecaption{Detectability results for all favorable targets at 12 pc using a GMT sized telescope. Missing values indicate that the molecular band was not detectable at the missing significance level in less than 300 transits. }\tablehead{
\colhead{Host Star} & \colhead{Atmosphere} & \colhead{Molecule} & \colhead{Band [$\mu$m]} & \colhead{\# transits to 3-$\sigmadet$} & \colhead{\# transits to 5-$\sigmadet$}
}
\startdata
M2V & ARE & \ce{CO2} & 1.59 & 17 & 54 \\
M2V & ARE & \ce{H2O} & 0.9  & 270 & \nodata \\
M2V & PIE & \ce{CH4} & 1.6 & 54 & \nodata \\
M2V & PIE & \ce{CO2} & 1.59 & 22 & 58 \\
M2V & PIE & \ce{H2O} & 0.9  & 150 & \nodata \\
M2V & PIE & \ce{O2} & 1.27  & 120 &  \nodata \\
M3V & ARE & \ce{CO2} & 1.59 & 16 &  62 \\
M3V & PIE & \ce{CH4} & 1.6 & 56 &  \nodata \\
M3V & PIE & \ce{CO2} & 1.59 & 23 &  62 \\
M3V & PIE & \ce{H2O} & 0.9  & 150  & \nodata \\
M3V & PIE & \ce{O2} & 1.27  & 190  & \nodata \\
M4V & ARE & \ce{CH4} & 1.6  & 54  & \nodata \\
M4V & ARE & \ce{CO2} & 1.59  & 16  & 60 \\
M4V & ARE & \ce{H2O} & 0.9  & 200  & \nodata \\
M4V & \ce{CO2} phot. ltng ON & \ce{CO2} & 1.59  & 18 &  60 \\
M4V & \ce{CO2} phot. ltng OFF & \ce{CO2} & 1.59  & 22  & 70 \\
M4V & \ce{CO2} phot. ltng OFF & \ce{CO} & 1.55  & 150  & \nodata \\
M4V & PIE & \ce{CH4} & 1.6 & 37 &  \nodata \\
M4V & PIE & \ce{CO2} & 1.59 & 23  & 60 \\
M4V & PIE & \ce{H2O} & 0.9 & 130  & \nodata \\
M4V & PIE & \ce{O2} & 1.27 & 150 &  \nodata \\
M6V & ARE & \ce{CH4} & 1.6 & 6 &  140 \\
M6V & ARE & \ce{CO2} & 1.59 & 21  & 62 \\
M6V & ARE & \ce{H2O} & 0.9 & 270  & \nodata \\
M6V & Ocean Loss & \ce{CO2} & 1.59 & 30 &  96 \\
M6V & Ocean Loss & \ce{O2} & 1.27 & 150 &  \nodata \\
M6V & Ocean Outgassing & \ce{CO2} & 1.59 & 29 &  140 \\
M6V & Ocean Outgassing & \ce{O2} & 1.27 & 110  & \nodata \\
M6V & PIE & \ce{CH4} & 1.6 & 8  & \nodata \\
M6V & PIE & \ce{CO2} & 1.59 & 20 &  60 \\
M6V & PIE & \ce{H2O} & 0.9 & 180 &  \nodata \\
M6V & PIE & \ce{O2} & 1.27 & 76  & \nodata \\
M8V & ARE & \ce{CH4} & 1.6 & 31 &  \nodata \\
M8V & ARE & \ce{CO2} & 1.59 & 26  & 290 \\
M8V & PIE & \ce{CH4} & 1.6 & 10 &  \nodata \\
M8V & PIE & \ce{CO2} & 1.59 & 30  & 240 \\
M8V & PIE & \ce{O2} & 1.27 & 58  & \nodata \\
\hline\enddata 
\end{deluxetable*}

\bibliography{high_res_bib}
\bibliographystyle{aasjournal}

\end{document}